# Theory and Applications of Infinitesimal Dipole Models for Computational Electromagnetics

Said M. Mikki and Ahmed A. Kishk
Center of Applied Electromagnetic Systems Research,
Department of Electrical Engineering, University of Mississippi,
University, MS 38677, USA

*Abstract* –**The recently introduced quantum particle swarm optimization (QPSO) algorithm is employed to find infinitesimal dipole models (IDM) for antennas with known near-fields (measured or computed). The IDM can predict accurately both the near-fields and the far- fields of the antenna. A theory is developed to explain the mechanism behind the IDM using the multipole expansion method. The IDM obtained from single frequency solutions is extrapolated over a frequency range around the design frequency. The method is demonstrated by analyzing conducting- and dielectric- type antennas. A calibration procedure is proposed to systematically implement infinitesimal dipoles within existing MOM codes. The interaction of the IDM with passive and active objects is studied through several examples. The IDM proved to predict the interaction efficiently. A closed-form expression for the mutual admittance between similar or dissimilar antennas, with arbitrary orientations and/or locations, is derived using the reaction theorem.**

## I. Introduction

Fast and efficient techniques to perform accurate predictions of EM devices are vital for the development of modern electronic telecommunication systems. As the complexity of various antenna systems and devices increases, satisfactory results can be obtained only by full-wave solutions. However, the cost of such solutions is very high and grows exponentially with the complexity and the size of the problem. To cope with these difficulties, alternative methods have been proposed. Such methods should be simple enough, and yet provide accurate predictions for desired quantities within a reasonable time.

A simplified EM method can be viewed as a technique for constructing a problem equivalent to the actual antenna under test (AUT). The model should correspond to the physical problem in the sense of producing similar electromagnetic fields when both the AUT and its equivalent model are interacting with the same environment.

Here, we use a set of infinitesimal dipoles to represent the AUT. This concept was first introduced to model the near-field of radiating structures in biomedical applications [1]. Later, the process was formulated as an optimization problem solved by evolutionary genetic algorithm (GA) techniques [2], [3]. In [4], the basic idea was refined and extended by applying the GA to general inverse source modeling problems using infinitesimal dipoles with both electric and magnetic types. However, the main purpose there was to predict the far-field performance of the AUT based on the measured near-field (amplitude only). The idea was further generalized by applying the GA to different antenna types, including dielectric resonator antennas (DRA) [5]. The GA, however, suffers from several drawbacks, such as the high computational complexity inherent in the method itself. It has been found in [4] and [5] that a model of eight dipoles represents an upper limit beyond which convergence problems start to degrade the performance of the technique. In order to model accurately antennas with rapid variations in the near-field distribution, larger number of dipoles is required. This can be conceptually justified if we interpret the extra dipoles needed as adding more degrees of freedom, allowing therefore for better accuracy in the field prediction.

Here, a newly introduced global optimization method, the quantum particle swarm optimization (QPSO) algorithm [6], is applied to tackle a dipole model version different from the one used in [4] and [5]. In former investigations of the applicability of the method to electromagnetic problems, the QPSO proved to be more efficient than its classical counterpart, the particle swarm optimization (PSO), in most of the examples tried [7]. In previous work using the GA, each dipole was associated with 10 degrees of freedom [4], [5]. However, in this article we introduce a simplified model that assigns only 7 degrees of freedom to each dipole and still achieve the same accuracy. While utilizing less number of degrees of freedom, the performance of the optimization using the QPSO proved to be better than previous GA solutions found in [5].

This article is organized as follow. First, the dipole moments and positions are obtained using the QPSO algorithm applied to the AUT. Comparisons between the actual near- and far- fields, together with their corresponding IDM's results, are presented. Second, a theoretical study is presented to analyze how and why the model works. Third, the performance of the model in providing a wideband prediction of the antenna characteristics, based on the center frequency model, is demonstrated. Then, a detailed study of the interaction of the AUT with some passive and active scatterers is given. A method for predicting the mutual coupling between two antenna elements based on the knowledge of the IDM of one element is proposed. Finally, conclusions are given.

## II. Formulation of the Problem

We start by an arbitrary antenna, contained in volume *V*, with near-field distribution $\left(\mathbf{E}^a, \mathbf{H}^a\right)$ observed over a surface *S* as shown in Fig.1. Assume a set of infinitesimal electric dipoles $\{\chi_i\}_{i=1}^N$, where *N* is the number of dipoles and $\chi_i$ is a seven-element vector representing the parameters of the *i*th dipole given by

$$\chi_i = \left[\mathrm{Re}\{M_i\} \quad \mathrm{Im}\{M_i\} \quad \alpha_i \quad \beta_i \quad x_i \quad y_i \quad z_i\right]^T \tag{1}$$

Here, the position of the *i*th dipole is given by $x_i$, $y_i$, and $z_i$, which are constrained by the actual antenna size. $M_i$ is the complex dipole moment, with orientation given by the direction cosines $\alpha_i$ and $\beta_i$ defined with respect to the *x*- and *y*- axis, respectively. The third directional cosine can be obtained from

$$\gamma_i^2 = 1 - \alpha_i^2 - \beta_i^2 \tag{2}$$

The components of the *i*th dipole moments are then given as

$$M_{ix} = M_i \alpha_i, \quad M_{iy} = M_i \beta_i, \quad M_{iz} = M_i \gamma_i \tag{3}$$

Notice that by employing directional cosines in the formulation, equations (2) and (3) eliminate two degrees of freedom from the total number associated with each dipole as used with the representation of [4] and [5]. Moreover, by restricting the dipole type to be electric only, we attain further reduction by one degree of freedom for each dipole. This gives seven variables per dipole and a total of 7*N* for the entire optimization problem.

By labeling the fields generated by the dipoles as $\left(\mathbf{E}^d, \mathbf{H}^d\right)$, we may define the cost function as follow

$$\begin{aligned} F = & w_e \left\{ \left( 1 \Big/ \sum_{n=1}^{N_{\text{ops}}} \sum_u \left| \mathbf{E}_u^a(\mathbf{r}_n) \right|^2 \right) \sum_{n=1}^{N_{\text{ops}}} \sum_u \left| \mathbf{E}_u^a(\mathbf{r}_n) - \mathbf{E}_u^d(\mathbf{r}_n) \right|^2 \right\}^{\frac{1}{2}} \\ & + w_h \left\{ \left( 1 \Big/ \sum_{n=1}^{N_{\text{ops}}} \sum_u \left| \mathbf{H}_u^a(\mathbf{r}_n) \right|^2 \right) \sum_{n=1}^{N_{\text{ops}}} \sum_u \left| \mathbf{H}_u^a(\mathbf{r}_n) - \mathbf{H}_u^d(\mathbf{r}_n) \right|^2 \right\}^{\frac{1}{2}} \end{aligned} \tag{4}$$

where $N_{\text{ops}}$ is the number of observation (samples) points of the near-field data set. The position vector of the *n*th sampling point is $\mathbf{r}_n$, where $\mathbf{r}_n \in S$ . The factors $w_e$ and $w_h$ are normalization coefficients in the range from 0 to 1; they represent weighing factors for the electric and magnetic field contributions to the objective function, respectively.

In general, this cost measure is highly nonlinear, with a landscape full of local minima. This makes the optimization problem very difficult unless a powerful global search method is used. In section IV, the recently introduced quantum particle swarm optimization (QPSO) [6], [7] is introduced to obtain the IDM.

The current distribution of the dipoles can be written as follow

$$\mathbf{J}_d(\mathbf{r}) = \sum_{n=1}^{N} M_n \delta(\mathbf{r} - \mathbf{r}_n') \left[ \hat{x} \cos\alpha_n + \hat{y} \cos\beta_n + \hat{z} \cos\gamma_n \right] \tag{5}$$

If the dyadic Green's function of the medium is given by $\bar{\mathbf{G}}_d^J(\mathbf{r}, \mathbf{r}')$, then the fields radiated by the antenna are [13]

$$\mathbf{E}_d(\mathbf{r}) = -j\omega\mu(\mathbf{r}) \iiint_V d\,\text{v}' \; \bar{\mathbf{G}}_d^J(\mathbf{r}, \mathbf{r}') \cdot \mathbf{J}_d(\mathbf{r}') \tag{6}$$

Substituting (5) into (6) we obtain

$$\mathbf{E}_d(\mathbf{r}) = -j\,\omega\,\mu(\mathbf{r}) \sum_{n=1}^{N} M_n \bar{\mathbf{G}}_d^J(\mathbf{r}, \mathbf{r}_n') \cdot \left[ \hat{x} \cos\alpha_n + \hat{y} \cos\beta_n + \hat{z} \cos\gamma_n \right] \tag{7}$$

Equation (7) can be interpreted as an approximation of the antenna's fields in terms of $N$ dipoles. It is interesting to notice that by finding the IDM of (7) we get a closed-form analytical solution for a complicated EM problem, which works everywhere outside the region of the antenna itself.

## III. Theory of Infinitesimal Dipole Models

The main effort of the first application of the GA to obtain IDMs focused on EMC problems arising from PCB devices. The theoretical justification of the method was basically empirical and related to the specific system under consideration [4]. In this section, we present a complete theory of how and why a set of infinitesimal dipoles can model the complicated problem of arbitrary antenna configurations.

### III.A) Analytic Continuation of the Electromagnetic Fields

First, let us see *how* knowledge of the near-fields over a finite region in space (surface $S$ in Fig. 1), can lead to full determination of the fields everywhere outside the antenna region (this region itself contains the dipoles). Notice that there are no restrictions imposed on $S$ to be either closed or open. However, even if the near-field observation surface is considered closed, the problem is inherently different from Huygens equivalence theorem since the equivalent sources we are searching for are not placed on a Huygens closed surface.

The appropriate analysis of the IDM relates to the concept of *non-radiating sources*, which arises in the generalized source equivalence theory [9]. If two sources radiate the same fields over a region in space, we call the two sources *equivalent*. According to the superposition theorem, it is obvious that the difference of these two sources is a non-radiating (NR) source since they radiate null-fields in the same region.

Regular regions are defined as those regions in space whose 1) media parameters are analytic functions and in which also 2) boundary surfaces, separating two different media, have continuous tangents [9]. According to [10], it can be shown that the electromagnetic fields, which are obtained upon solving a special form of partial differential equations dictated by Maxwell's equations, are analytic functions of position. Therefore, by using analytic continuation, if a zero-field is synthesized in the sub-domain $S \subset V$ , where $V$ is the domain of analyticity, then the fields are also zero everywhere else in $V$. This means that if we can find an equivalent source that produces the same fields on a surface $S$ contained within a larger regular region $V$, then the source (now a model) can predict the fields everywhere in $V$.

Now, one might ask how small the surface $S$ should be? In general, the theory above does not place any restriction on the size of the observation surface. However, in practice, as will be demonstrated in the coming sections, appropriate choices for the size of this surface and its distance from the antenna turn out to be critical. The reason is related to the sampling theorem. The analytical continuation requires that the fields of NR sources are exactly zero *everywhere* over $S$. This leads to infinite number of degrees of freedom since the surface is a continuous domain. In reality, one deals only with finite number of samples; however, if the sampling theorem is satisfied, "enough" samples can generate full knowledge of the field behavior over the continuous domain. How much "enough" points are taken depends largely on the nature of the field variations over the observation plane, which in turns determines the size of the plane and its distance from the antenna.

### III.B) Multipole Expansion Theory for Infinitesimal Dipole Models

In order to explain *why* a set of point sources (or dipoles) are capable of representing the AUT, we will use the theory of multipole expansion of sources to attain better understandings of the problem. Depending on the type of the AUT, we can always employ either surface or volume equivalence theorems to replace the antenna by equivalent surface or volume currents, respectively. Without loss of generality, suppose we have a volume-type antenna. Application of volume equivalence theorem leads to equivalent volume current density $\mathbf{J}(\mathbf{r})$ contained within a volume $V$ [12].

Assume that there exists a small region $D_n \subset V$ . Using multipole theory, we can write the expansion of a source located within that volume element as [9]

$$\mathbf{J}_{D_n}(\mathbf{r}) = \delta(\mathbf{r}-\mathbf{a}_n)P_0^n - \nabla\delta(\mathbf{r}-\mathbf{a}_n)\cdot P_1^n(\mathbf{a}_n) + \nabla\nabla\delta(\mathbf{r}-\mathbf{a}_n) : P_2^n(\mathbf{a}_n) - \dots$$
$$= \sum_{m=0}^{\infty} \frac{(-1)^m}{m!} \underbrace{\nabla\nabla\dots\nabla}_{m \text{ times}} \delta(\mathbf{r}-\mathbf{a}_n) \overset{m}{\odot} P_m^n(\mathbf{a}_n) \quad (8)$$

where

$$P_m^n(\mathbf{a}_n) = \iiint_{D_n} d\,\mathbf{r}'^3 \underbrace{(\mathbf{r}'-\mathbf{a}_n)(\mathbf{r}'-\mathbf{a}_n)\dots(\mathbf{r}'-\mathbf{a}_n)}_{m \text{ times}} \mathbf{J}_{D_n}(\mathbf{r}') \quad (9)$$

is the $m$th moment of the source in the region $D_n$ (polyad of rank $m$). The operation $\overset{n}{\odot}$ stands for $n$-dot product between two $n$-ads (polyads of rank $n$) and $\delta$ stands for the Dirac delta function with its higher-order derivatives are understood in the sense of generalized function theory. In the expansion above, $\mathbf{a}_n$, which is possibly complex, refers to the location of the point where the multipole expansion is performed.

If it is desired to directly represent the current distribution $\mathbf{J}(\mathbf{r})$ in terms of point sources, as illustrated in Fig.2 (a), then one must according to the sampling theorem use a large number of sources covering the whole volume $V$. This is because the multipole expansion in (8) can lead to point sources (i.e., the first term in the series is the dominant) only if the sources themselves are highly localized in space. It is obvious, however, that using a large number of sources in the dipole model is not practical; this is because the computational demands in the optimization problem, and the complexity of the fitness landscape, will make the solution very difficult, if not impossible.

Fig.2 (b) illustrates another case where few sources are allowed to represent the volume current $\mathbf{J}(\mathbf{r})$. Here each of the $N$ sources exists within a *finite* volume $D_n$ such that

$$V = \bigcup_{n=1}^{N} D_n \quad (10)$$

For simplicity, only four sources are illustrated in Fig. 2 (b). Employing the expansion in (8) to represent each of these sources, the total current expansion can be written as

$$\mathbf{J}^a(\mathbf{r}) = \sum_{n=1}^{N} \mathbf{J}_n^a(\mathbf{r}) = \sum_{n=1}^{N} \sum_{m=0}^{\infty} \frac{(-1)^m}{m!} \underbrace{\nabla\nabla\ldots\nabla}_{m \text{ times}} \delta(\mathbf{r}-\mathbf{a}_n) \overset{m}{\odot} P_m^n(\mathbf{a}_n) \tag{11}$$

If we separate the first-order terms from the higher-order terms, we can re-write equation (11) as

$$\mathbf{J}^a(\mathbf{r}) = \underbrace{\sum_{n=1}^{N} \delta(\mathbf{r}-\mathbf{a}_n) P_0^n(\mathbf{a}_n)}_{\text{First-order terms}} + \underbrace{\sum_{n=1}^{N} \sum_{m=1}^{\infty} \frac{(-1)^m}{m!} \underbrace{\nabla\nabla\ldots\nabla}_{m \text{ times}} \delta(\mathbf{r}-\mathbf{a}_n) \overset{m}{\odot} P_m^n(\mathbf{a}_n)}_{\text{Higher-order terms}} \tag{12}$$

In light of equation (12), the optimization problem can be re-formulated as follow: *Search for the set of source volume localizations, specified by* $\{D_n\}_{n=1}^{N}$, *the location vectors, specified by* $\{\mathbf{a}_n\}_{n=1}^{N}$, *and current distributions, specified by* $\{\mathbf{J}_{D_n}\}_{n=1}^{N}$, *such that the second double summation in (12) is zero. In other words, all of the higher-order terms of the N sources should cancel out.* One may express this search process as a minimization of the following cost function

$$F' = \left\| \sum_{n=1}^{N} \sum_{m=1}^{\infty} \frac{(-1)^m}{m!} \underbrace{\nabla\nabla\ldots\nabla}_{m \text{ times}} \delta(\mathbf{r}-\mathbf{a}_n) \overset{m}{\odot} P_m^n(\mathbf{a}_n) \right\|^2 \tag{13}$$

where $\| \;\|$ indicates the norm defined as $\|\mathbf{A}\| = \sqrt{\mathbf{A}\cdot\mathbf{A}^*}$. By dealing with this optimization problem, instead of (4), the solution process becomes obviously more difficult. However, the utilization of the theorem stated in Section A indicates that the solutions obtained in (4) and (13) are equivalent. It is clear that the formulation of (4) is much easier to be implemented in a general-purpose code, while the one in (13) provides us with deeper insight into the nature of the problem. For example, by considering the form of (12), it is possible to see that the dipole model described in (5) is indeed a first-order approximation of the original problem in the multipole expansion sense. Notice that according to (9) $P_0^n$ is *not* a function of $\mathbf{a}_n$. Therefore, when the overall contribution of the higher-order terms is zero, $P_0^n$ in (12) and $\mathbf{M}_n$ in (5) are identical.

The multipole theory assumes that the source series expansion and the original source are equivalent only over the volume outside the source region *V*. This means that the IDM obtained is valid only in this region. On the other hand, since arbitrary number of NR sources can be added to the series expansion (8) without affecting the total field, then multiple possible multipole expansion can be obtained. Consequently, there is no unique solution for the optimization problem in (13). This was observed in our actual implementation of the method using the evolutionary QPSO algorithm, where different solutions were obtained through several runs.

## IV. Application of the QPSO Algorithm to Find Infinitesimal Dipole Models

### IV.A) Boundary Condition for the Direction Cosines

In [7], the boundary condition (BC) that was used to truncate the particle's position in the QPSO algorithm is the hard domain boundary condition [11]. However, because the two direction cosines, as defined in (1), are coupled to each other, the direct application of this type of BC becomes not possible. To overcome this difficulty, a new BC is proposed. Fig. 3 illustrates how the BC for the two directional cosines $\alpha$ and $\beta$ should be chosen. Every time a particle falls outside the unit circle, we move it to the boundary of the circle while keeping its angle with $\alpha$ fixed. This will guarantee that none of the particles are allowed to fall outside the physical domain of search.

### IV.B) Infinitesimal Dipole Models for Conducting and Dielectric Antennas

In order to illustrate the applicability of the method outlined above, the QDPSO version of the QPSO algorithm in [7] is used to find IDMs for two practical antenna configurations: Conducting patch antenna and dielectric resonator antenna (DRA). A conducting patch excited by L-shaped coaxial probe is shown in Fig. 4. The patch is designed to resonate at 4.5 GHz with a matching bandwidth about 27%. An accurate Method of Moment (MoM) solution [8] is obtained. The near-field is computed on a square observation plane, with side lengths $5\lambda$, located at a distance $\lambda$ from the ground plane. The total number of samples is 2025 points. The geometry of a circular DRA located above an infinite PEC ground plane is shown in Fig. 5, where a coaxial probe excites the antenna. The matching bandwidth has a center frequency of 10 GHz. The MoM procedure is used to analyze the structure. Near-field data are computed at a square plane of side lengths $\lambda$. The distance of the observation plane from the ground is taken to be $\lambda$. The total number of samples is 625 points.

It should be mentioned that compared to the planar near-field measurement procedure performed usually in the laboratory, the near-field plane is located at several wavelengths from the antenna. Such experimental procedures are valid for high directive antennas and predict the far-fields only within a small view range, which is a function of the near-field plane size. However, the present method can predict the far-field for a considerably wider observation view range that could be the entire space. Here, the sampling rate is denser than those in the planar near-field measurements. One can consider this near-field as another equivalent problem with at least 10 samples per wavelength as the required sampling rates in numerical methods (e. g., MoM). This allows using smaller number of dipoles to simulate the antenna performance. Each dipole or couple of dipoles may represent a mode that is excited in the antenna. If we are interested just in the far-field, only two dipoles with pre-specified polarization are sufficient to model the antenna. It is interesting to notice that by having arbitrarily oriented dipoles, and moving the near-field plane further away from the dipoles, it is possible to compare the IDM with the method of auxiliary sources (MAS) [14]. However, in MAS, the number of point sources located on the auxiliary sources is much higher compared with the IDM presented here. This is a direct consequence of allowing the point sources to move freely in a volume containing the antenna, instead of restricting these sources to pre-specified polarization and position, which on a surface conformal to the physical surface of the antenna.

The QDPSO algorithm is used to minimize the objective function defined in (4) by searching for electric dipoles with $w_e = 1$ and $w_h = 0$. The frequency of the IDM is identical to the design frequency of each antenna (4.5 GHz for the conducting patch and 10GHz for the DRA). The dipoles locations are restricted to be around the physical domain of the antenna itself, as demonstrated in Section III.B. The dipole moments are chosen based on the order of magnitude of the near-field data; therefore, the moments' search range should produce fields that are at the same order of magnitude as compared to the actual near-field data. In this example the ranges used for the DRA and the conducting patch are $\pm 1.0\times 10^{-5}$ A.m and $\pm 1.0\times 10^{-3}$ A.m, respectively. It should be pointed out that if an inappropriate search space for the dipoles' locations and moments is chosen, then regardless of the number of dipoles, the number of iterations, and population size used, the method may not converge since no physical solution exists.

In order to improve the convergence of the method and reduce the number of unknowns, image theory is used when the antenna is above an infinite ground plane. Thus, each dipole is automatically associated with its image. This will also assure that the obtained dipoles represents the antenna itself without the ground plane, and therefore can be used to study the interaction with a finite ground plane as will be shown later. The accuracy of the solution is expressed in terms of the global error defined as

$$e = \frac{1}{2}\frac{\left|\sum_{n=1}^{N_{\text{ops}}}\sum_{u}\left(E_{n,u}^{\text{actual}} - E_{n,u}^{\text{dipoles}}\right)\right|}{\sum_{n=1}^{N_{\text{ops}}}\sum_{u}\left|E_{n,u}^{\text{actual}}\right|} + \frac{1}{2}\frac{\left|\sum_{n=1}^{N_{\text{ops}}}\sum_{u}\left(H_{n,u}^{\text{actual}} - H_{n,u}^{\text{dipoles}}\right)\right|}{\sum_{n=1}^{N_{\text{ops}}}\sum_{u}\left|H_{n,u}^{\text{actual}}\right|} \tag{17}$$

Two sets of 10 and 5 dipoles are considered for the patch antenna and the DRA, respectively. The QDPSO algorithm, with population size of 80 particles and 5,000 generations, was successful in obtaining good IDMs for the two antennas. The achieved global errors are 0.811% and 0.162%, for the conducting patch and the DRA respectively. Fig. 6 illustrates the convergence curve for the conducting patch. In Fig. 7, samples of the conducting patch antenna IDM's near-fields are compared with the MoM near-fields. The far-field radiation patterns of the IDM are computed using the analytical expressions given in [12] and compared with the MoM solution as shown in Fig. 8. Excellent agreement is found in both E- and H- planes.

To further test the model, the near-fields of the antenna were calculated at numerous plane cuts, with various angles, using both the MoM and the IDM. Excellent agreement was observed all the time. Although the IDM was obtained based on knowledge of the fields over a finite open observation plane, the model is capable of predicting the fields everywhere in the near- and far- zones within the global error of the model. Discrepancies, however, start to appear if the observation points are located very close to the dipole region. This is because the singularities of the infinitesimal sources will disturb the smoothness of the original field, causing the IDM prediction to deviate from the actual fields. This difficulty can be overcome by increasing the number of dipoles, which in return increases the optimization CPU time. However, this might be needed only if the IDM is used to study interactions with very close nearby objects. One of the advantages of the application of the QPSO algorithm to find the IDM is that it becomes possible to increase the number of the optimization variables beyond what was reported as an upper bound for the GA optimization method [4].

## V. The Frequency Response of Infinitesimal Dipole Models

### V.A) Motivation

As was indicated above, the IDM is obtained from single frequency measurements. In order for the model to be valid at various frequencies, additional measurements are required at these frequencies, and the previous optimization process has to be repeated at each frequency. In general, this will generate different sets of dipoles for each frequency. To avoid such lengthy process, one may fix the dipoles' orientations and find different dipole moments and positions. In some cases, the positions of the dipoles may be scaled by the frequency ratio and only the moments should be found by utilizing the optimization procedure, but for smaller number of variables or parameters. However, all these methods require either field measurements or full-wave solution performed over the frequency band of interest, which is generally an expensive process.

Here, we propose another method to obtain IDM models that are valid for a considerable frequency band. The method is based on observing the nature of the deviation between the fields generated by the IDM obtained at a single frequency, and those of the actual problem. Intuitively, it is expected that a "correction factor" for the dipole moments can accommodate for the frequency response within a range in which the field distribution will have small variations around the central frequency response. In other words, the antenna has to be operated within the frequency range of the same mode(s). If new mode is excited at other frequencies, the signature of the field distribution changes, which makes the dipole model not valid for these frequencies. To use the same IDM, one may employ a frequency-dependent complex factor to scale the dipole moments such that predication of the near- and far-fields, within certain frequency range, can be achieved. This method does not require changes in the positions and orientations of the dipoles once they are obtained through the optimization at the center frequency. Only the moments are required to vary slowly according to the correction factor provided with the model. Moreover, the correction factor is related to the input impedance of the AUT, which means that at each frequency one value is enough; this gives the proposed method certain advantage over the near-field measurement technique, which is time consuming and requires a large storage for the data. A simple mathematical proof for the idea presented above is given in the next section.

### V.B) Theory

Assume that the AUT is excited by a source represented by the volume current density $\mathbf{J}_i(\mathbf{r}')$. Let the dyadic Green's functions of the medium around the source be $\bar{\mathbf{G}}_i^J(\mathbf{r},\mathbf{r}';\omega)$, which is evaluated at the radian frequency ω. By equating the fields radiated by both the IDM and the actual antenna we may write

$$\iiint_V dv'\,\bar{\mathbf{G}}_d^J(\mathbf{r},\mathbf{r}';\omega)\cdot\mathbf{J}_d(\mathbf{r}';\,\omega)=\iiint_V dv'\,\bar{\mathbf{G}}_i^J(\mathbf{r},\mathbf{r}';\omega)\cdot\mathbf{J}_i(\mathbf{r}';\omega) \tag{18}$$

If the voltage excitation complex amplitude is kept constant, the variation in the excitation current will be inversely proportional to the input impedance $Z_{in}(\omega)$. Thus, we may express the excitation current at a new frequency $\omega$ as

$$\mathbf{J}_i(\omega)=\mathbf{J}_i(\omega_0)\frac{Z_{in}(\omega_0)}{Z_{in}(\omega)} \tag{19}$$

where $\omega_0$ is the center frequency. Also, the dipoles' currents at the new frequency will be assumed to obey the following relation

$$\mathbf{J}_d(\omega)=\mathbf{J}_d(\omega_0)\Upsilon(\omega) \tag{20}$$

where $\Upsilon(\omega)$ is a dimensionless complex unknown. Now, let the dyadic Green's functions of the dipole medium and the actual source medium be perturbed around the design frequency $\omega_0$ as follow

$$\bar{\mathbf{G}}_d^J(\mathbf{r},\mathbf{r}';\omega)=\bar{\mathbf{G}}_d^J(\mathbf{r},\mathbf{r}';\omega_0)+\delta\bar{\mathbf{G}}_d^J(\mathbf{r},\mathbf{r}';\omega) \tag{21}$$

$$\bar{\mathbf{G}}_i^J(\mathbf{r},\mathbf{r}';\omega)=\bar{\mathbf{G}}_i^J(\mathbf{r},\mathbf{r}';\omega_0)+\delta\bar{\mathbf{G}}_i^J(\mathbf{r},\mathbf{r}';\omega) \tag{22}$$

If the Green's functions capture the same modes within the frequency perturbation band, then it is reasonable to assume that these functions are slowly varying with frequency; thus, we may drop the terms $\delta\bar{\mathbf{G}}_d^J(\mathbf{r},\mathbf{r}';\omega)$ and $\delta\bar{\mathbf{G}}_i^J(\mathbf{r},\mathbf{r}';\omega)$ from equations (21) and (22) and approximate the Green's functions in the kernels of the integrations in (18) by their values at $\omega_0$ as follow

$$\bar{\mathbf{G}}_d^J(\mathbf{r},\mathbf{r}';\omega)\simeq\bar{\mathbf{G}}_d^J(\mathbf{r},\mathbf{r}';\omega_0) \tag{23}$$

$$\bar{\mathbf{G}}_i^J(\mathbf{r},\mathbf{r}';\omega)\simeq\bar{\mathbf{G}}_i^J(\mathbf{r},\mathbf{r}';\omega_0) \tag{24}$$

Substituting (19), (20), (23), and (24) into (18) we get

$$\iiint_V d v' \bar{\mathbf{G}}_d^J(\mathbf{r},\mathbf{r}';\omega_0)\cdot\mathbf{J}_d(\omega_0)\,\Upsilon(\omega)=\iiint_V d v' \bar{\mathbf{G}}_i^J(\mathbf{r},\mathbf{r}';\omega_0)\cdot\mathbf{J}_i(\omega_0)\frac{Z_{in}(\omega_0)}{Z_{in}(\omega)} \tag{25}$$

However, at the design frequency we know that

$$\iiint_V d v' \bar{\mathbf{G}}_d^J(\mathbf{r},\mathbf{r}';\omega_0)\cdot\mathbf{J}_d(\omega_0)\approx\iiint_V d v' \bar{\mathbf{G}}_i^J(\mathbf{r},\mathbf{r}';\omega_0)\cdot\mathbf{J}_i(\omega_0) \tag{26}$$

From (26) and (25) we find

$$\Upsilon(\omega)=\frac{Z_{in}(\omega_0)}{Z_{in}(\omega)} \tag{27}$$

which states that the required correction needed to be imposed over the dipole moments is related to the input impedance of the original antenna.

**V.C) Practical Implantations and Results**

Assume that an IDM was obtained at the center frequency $\omega_0$. We define the *relative* correction factor needed to predict the moments at a new frequency $\omega$ as

$$\gamma(\omega)=\frac{M_u(\omega)}{M_u(\omega_0)} \tag{28}$$

where *u = x, y, z*, $M_u(\omega)$ is the *u*-component of dipole moment at the frequency under consideration ω, and $M_u(\omega_0)$ is the moment at the design frequency $\omega_0$. One motivation behind this definition is that the moments appear in the expressions of the radiated fields of the infinitesimal dipoles as multiplicative factors [12]. Thus, if the variation of the moments with frequency can be described by a single scaling factor, then the overall change in the electromagnetic field can be described by the same factor. In other words, for the IDM to be valid at several frequencies, we do not need to run the optimization process again to obtain different dipoles at different locations. We need only to change the moments by a correction factor that is valid for all the dipoles. To obtain this factor, $\gamma$, get the ratio of the fields at the new frequency over the fields at the design frequency as follow

$$\gamma(\omega)=\frac{\Psi(\omega)}{\Psi(\omega_0)} \tag{29}$$

where $\Psi$ indicates *averaged* values for either the electric or the magnetic fields components. The accuracy of this estimation can be enhanced if six correction factors are obtained in the following way

$$\gamma_u^{\Psi}(\omega) = \frac{1}{N_{\text{eff}}} \sum_{n=1}^{N_{\text{ops}}} \Psi_u(\mathbf{r}_n;\omega) W(\mathbf{r}_n) \tag{30}$$

where $\Psi$ is either **E** or **H**. Here, $W(\mathbf{r}_n)$ is a window function defined as

$$W(\mathbf{r}_n) = \begin{cases} \dfrac{1}{\Psi_u(\mathbf{r}_n, \omega_0)}, & \Psi_u(\mathbf{r}_n, \omega_0) \neq 0 \\ 0, & \Psi_u(\mathbf{r}_n, \omega_0) = 0 \end{cases} \tag{31}$$

and is used to avoid calculating the ratio when the denominator is zero. $N_{\text{eff}}$ is the actual number of samples considered in the calculations of (30). The original correction factor in (29) can be estimated as the average of the six correction factors obtained in (30). As has been found in (27), the following relation also holds

$$\gamma(\omega) = \frac{Z_{in}(\omega_0)}{Z_{in}(\omega)} \tag{32}$$

Equations (30) or (32) provide two different methods to estimate the relative correction factor $\gamma$ at several discrete frequencies. Later, a suitable interpolation law may be imposed to construct the full curve through the intermediate frequencies. This will lead to accurate estimation for $\gamma$ within the required bandwidth.

Fig. 9 shows a comparison between the relative correction factors calculated using both equations (30) and (32) for the conducting patch antenna. In Fig. 10, the frequency responses of both the conducting patch and the DRA are shown. The global error in (17) is used to measure the accuracy of the overall performance of the model at various frequencies. The results in Fig. 7 show the fields at the design frequency within the modes bandwidth. In order to help the reader in appreciating the amount of error in Fig. 10, we show in Fig. 11 the conducting patch near-field comparison corresponding to a global error of 16.1%. It is clear that at such an error level, the IDM is not valid for this frequency; the near-field results indicated that there is another mode excited that was not present at the central frequency.

It is important to mention that the IDM's bandwidth should be limited to the bandwidth of the modes excited at the center frequency as previously explained. Far from this frequency we expect other modes to contribute; however, their contributions were not considered in the obtained IDM. It is interesting to observe how the error peaks when the frequency is decreased and then drops rapidly in the very low frequency range. The above results indicates that the IDM achieves a percentage bandwidth around 200% for the conducting patch, and 190% for the DRA.

## VI. Implementation of IDMs within Existing MOM Codes

Most commercial software packages do not offer special treatment for ideal infinitesimal dipoles. In this section, we obtain a dipole model that can be implemented in commercial codes. One of these codes is WIPL-D, which is a full-wave MoM solver that can efficiently handle and model arbitrary-shaped dielectric and metallic objects [8].

We start by defining the dipole moment $M$ as

$$M = I\, l \tag{33}$$

where $I$ is the *uniform* current flowing through a dipole with length $l$. In WIPL-D, only voltage source excitations are allowed to excite a wire antenna. For a given wire length and radius, we must choose the suitable voltage source such that the current flowing through the wire will produce the same moment of the corresponding infinitesimal dipole in (33).

Consider a small wire as in Fig. 12. The length and the radius of the wire are taken to be $0.001\lambda$ and $0.0001\lambda$, respectively. From equation (33) we can write

$$V_n = \frac{Z_n^d}{l} M \tag{34}$$

where $V_n$ and $Z_n^d$ are the excitation voltage and the input impedance of the $n$th real dipole, respectively. Equation (34) implies that for a given dipole moment $M$, the voltage excitation can be found if the input impedance of the actual wire is known. However, such information about the input impedance is not always available with the required accuracy, which makes the direct application of (34) difficult.

In this section, a systematic calibration procedure is proposed to tune WIPL-D's dipoles by determining the vector of excitations $\{V_n\}_{n=1}^{N}$ for the IDM's $N$ dipoles. Based on (34), we assume that the relation between the voltage $V$ and the dipole moment $M$ is given by

$$V = C_F\, M \tag{35}$$

where $C_F$ is a complex number, called here the *calibration constant*, which can be found by applying the following procedure

1. Start by setting $V$ = 1.0 Volt for a very small wire dipole of known radius, length, and some arbitrary orientation, the $z$-direction for example.

2. Consider a single observation point located at a distance $\lambda$ normal to the wire axis. Use WIPL-D to calculate the fields generated by the dipole. Calculate the fields generated by an infinitesimal dipole with moment **M,** oriented in the same direction of WIPLD's dipole, using the analytical expressions. The calibration constant of the wire is given by

$$C_F = \frac{\Psi_{\text{Dipoles}}(\mathbf{M})}{M\ \Psi_{\text{WIPL-D}}(V = 1.0)} \tag{36}$$

where $\Psi$ is one of the six components of **E** and **H**. This constant is dependent only on the wire radius, length, and the surrounding medium properties.

3. Apply equation (35) to get the corresponding voltage excitation for a desired moment.

Based on this procedure, the near- and far- fields radiated by both of WIPL-D's realization of the IDM are compared against each other. Excellent agreement is observed.

## VII. Interaction with Passive and Active Environments

In this section, we will study the interaction with nearby objects or with other antennas when the AUT is replaced by its IDM. By applying the calibration procedure of section VI, it is possible to combine available MoM codes with the IDM to solve the equivalent problem. It should be noticed that we ignore the effect of the object on the antenna itself. If the original current distribution on the object is not affected much by the replacement of the AUT by its IDM, then we expect that good agreements between the new simplified problem and the original one can be achieved.

### VII.A) PEC Scatterers

First, we consider a DRA located above an infinite ground plane in the presence of a square conducting plate of dimensions $0.2\lambda \times 0.2\lambda$ placed above the DRA by a distance of 0.8λ as illustrated in Fig. 13 (a). The DRA is replaced by its equivalent set of dipoles as shown in Fig. 13(b). The far-fields produced by the actual antenna are compared to the fields obtained from the IDM in the presence of the conducting plate as shown in Fig. 14. Very good agreement between the two results is observed, which indicates that the IDM's interaction with PEC scatterers is a valid approximation.

The other example consists of a rotated PEC scatterer of dimensions $0.25\lambda \times 0.25\lambda$ located $0.75\lambda$ away from the conducting patch antenna as shown in Fig. 15(a). The dimensions of the patch itself are $0.21\lambda \times 0.23\lambda$, and are the same as the geometry depicted in Fig. 4. The IDM was used to predict the far-field in the presence of the scatterer and good agreement is reported in Fig. 15(b).

### VII.B) Dielectric Scatterers

The interaction with dielectric objects is also considered. Fig. 16 illustrates the geometry of a dielectric cube of dimensions $0.13\lambda \times 0.13\lambda \times 0.2\lambda$ and $\varepsilon_r = 3.0$ located at a distance $0.8\lambda$ above the DRA. The antenna itself is residing above a *finite* ground plane with dimensions $\lambda \times \lambda$. Notice that we introduce here two objects: the finite ground plane and the dielectric cube. Considering a finite ground plane is similar to studying this problem using geometrical theory of diffraction (GTD). However, the ground plane size used here is very small. As mentioned before, this method allows prediction of the far-field in a very wide view range (in this case the whole space, unlike in the near-field measurement that has limited far-field view range that depends on the near-field plane position and size). The comparison between the far-field radiations of the DRA and its IDM is shown in Fig. 17. Good agreement is observed.

Fig. 18(a) shows the geometry of a larger dielectric sphere with $\varepsilon_r = 5.0$ and diameter $0.3\lambda$ located $0.7\lambda$ away from the infinite ground plane. The far-field comparison is shown in Fig. 18(b) where good agreement is observed in the E- and H-planes.

### VII.C) Antenna-Antenna Interactions

In Fig 19, we show another example illustrating the use of IDMs to predict the interaction with active elements. Fig. 19(a) illustrates the geometry of E-plane 2-element DRA array. Fig. 19 (b) and (c) show the far-field comparison for an inter-element spacing of $\lambda$ and $0.67\lambda$, respectively. Very good agreement is observed.

### VII.D) Reduction in the Computational Demands

The computational cost of using the IDM are obviously much less compared with the direct full-wave solution of the original problem. Although a full-wave procedure is still required with the simplified version, the MOM solution of the original problem discretized the antenna surface, which increases the total number of unknown. Therefore, the utilization of the IDM leads to an elimination of almost all of the unknowns that were used by the MOM to model the antenna itself. The numbers of unknowns required in WIPL-D solutions for the various examples tried in this section are listed in Table I.

**Table I** Number of unknowns required for the MOM solution of antenna interactions with passive and active elements

| Physical Problem/Model used | No. of unknowns in the original problem | No. of unknowns when IDM is used |
|---|---|---|
| DRA with PEC square scatterer | 318 | 9 |
| DRA with dielectric cube scatterer | 954 | 373 |
| Patch with PEC square scatterer | 140 | 34 |
| Patch with spherical dielectric scatterer | 1564 | 394 |
| DRA-DRA interaction | 612 | 311 |

## VIII. Simple and Efficient Method to Calculate the Mutual Coupling between Antenna Arrays Elements

Mutual coupling in antenna arrays refers to the induction of currents in one element by the radiated fields of another element. In designing arrays, it is important to study mutual coupling between each two elements with the presence of the remaining elements. If the mutual interaction is strong, then this may affect the radiation pattern, directivity, and the input impedance of the elements. However, attaining accurate estimations for the mutual coupling requires full-wave analysis of the array, which is very expensive in general.

In this section, a simple and efficient method to calculate the mutual coupling between two arbitrary antennas is proposed by employing the IDM that was obtained for the single element configuration. The reaction theorem is applied to derive an analytical expression for the mutual admittance of the two antennas (not necessary be identical to each other).

Fig. 20 illustrates the geometry of the problem where two arbitrary antennas are considered. Each antenna may be replaced by its IDM, as shown in the figure. The electric and magnetic currents $(\mathbf{J}_1, \mathbf{M}_1)$ and $(\mathbf{J}_2, \mathbf{M}_2)$ will produce electromagnetic fields $(\mathbf{E}_1, \mathbf{H}_1)$ and $(\mathbf{E}_2, \mathbf{H}_2)$ for antennas 1 and 2, respectively. We start by the following expression for the mutual admittance that can be easily derived from the reaction theorem [12]

$$Y_{12} = \frac{1}{V_1 V_2} \iiint_{V'} \left[ \mathbf{E}_1 \cdot \mathbf{J}_2 - \mathbf{H}_1 \cdot \mathbf{M}_2 \right] dv' \tag{38}$$

where $V_1$ and $V_2$ are the ports voltages. $(\mathbf{E}_1, \mathbf{H}_1)$ and $(\mathbf{E}_2, \mathbf{H}_2)$ are the fields radiated by the first and second antennas, respectively. $V'$ is the volume of integration that includes one of the two antennas' physical domains. For a set of electric dipoles, the current distribution in the second antenna is given by

$$\mathbf{J}_2(\mathbf{r}, \mathbf{r}') = \sum_{n=1}^{N} \mathbf{M}_n^J \delta(\mathbf{r} - \mathbf{r}_n') \tag{39}$$

where $N$ is the number of dipoles. $\mathbf{r}_n'$ and $\mathbf{M}_n^J$ are the location and the moments of the $n$th electric dipole. By substituting (39) into (38), and using the fact that $\mathbf{M}_2 = 0$, we find

$$Y_{12} = \frac{1}{V_1 V_2} \sum_{n=1}^{N} \mathbf{E}_d(\mathbf{r}_n') \cdot \mathbf{M}_n^J \tag{40}$$

where $\mathbf{E}_d(\mathbf{r}_n')$ is the field of the other antenna (in this case $\mathbf{E}_1$) evaluated at the dipole's positions $\mathbf{r}_n'$. In light of equation (5) we can finally write

$$Y_{12}(f) = \frac{1}{V_1 V_2} \sum_{n=1}^{N} \mathbf{E}_d(\mathbf{r}_n'; f) \cdot M_n^J \, \Upsilon(f) \left[ \hat{x} \cos\alpha_n + \hat{y} \cos\beta_n + \hat{z} \cos\gamma_n \right] \tag{41}$$

If the fields $\mathbf{E}_d(\mathbf{r}_n')$ are not available through measurements or numerical solution, it is possible to estimate them by employing the wideband IDM itself. Thus, we finally obtain

$$Y_{12}(f) = \frac{-j\omega\mu\Upsilon(f)^2}{V_1 V_2} \sum_{m=1}^{N}\sum_{n=1}^{N} M_n^J M_m^J \left[ \bar{\mathbf{G}}_d^J(\mathbf{r}, \mathbf{r}'_n; f) \cdot \hat{\mathrm{I}}_n \right] \cdot \hat{\mathrm{I}}_m \tag{42}$$

where

$$\hat{\mathrm{I}}_n = \hat{x}\cos\alpha_n + \hat{y}\cos\beta_n + \hat{z}\cos\gamma_n \tag{43}$$

Equation (42) represents the mutual coupling between two antennas presented in a simple closed-form expression.

As an example, we consider first E-plane two DRA elements. Fig. 21 illustrates the comparison between the mutual admittance calculated using our method and the MOM solution obtained using WIPL-D. It is clear that the proposed method, in spite of its simplicity, predicts the mutual coupling between the two elements for a considerable frequency band.

The mutual coupling between two conducting patches in the E-plane is also considered. Fig. 22 illustrates the comparison between the MOM solution and the present method. Good agreement is observed.

## IX. Conclusion

In this paper, a systematic method to model arbitrary antennas by infinitesimal dipole models (IDM) was introduced. The QPSO algorithm was applied to find a model that predicts accurately the near- and far- fields of the antenna under test (AUT). A theory was proposed using the multipole expansion technique to justify the use of the model. It was found that the IDM represents a source equivalence characterization of the AUT different from traditional Huygens or volume equivalence theorems.

The method was used to provide IDMs for two different antennas, namely a conducting patch excited by L-probe and circular DRA excited by a coaxial probe. Good convergence results for the QPSO algorithm were reported, which outperform previous works with the GA. The IDM produced accurate predictions of the actual fields radiated by the antenna.

The IDM was found to exhibit a wideband performance when a frequency dependent correction factor was introduced to adjust the dipole moments. A calibration procedure was proposed to implement the obtained IDMs in existing MoM solvers. The IDM was used with the MoM to study the interaction of the AUT with nearby passive and active objects. The proposed method provided a computationally very efficient method to predict the interaction of an antenna with other objects without performing full-wave solution. A closed-form expression for the mutual coupling between two antenna elements was derived using the reaction theorem. The derived expression provided very good prediction for the mutual coupling of the two antennas considered above. The presented work could be used to create a library of IDM for many antennas of practical interest for antenna engineers.

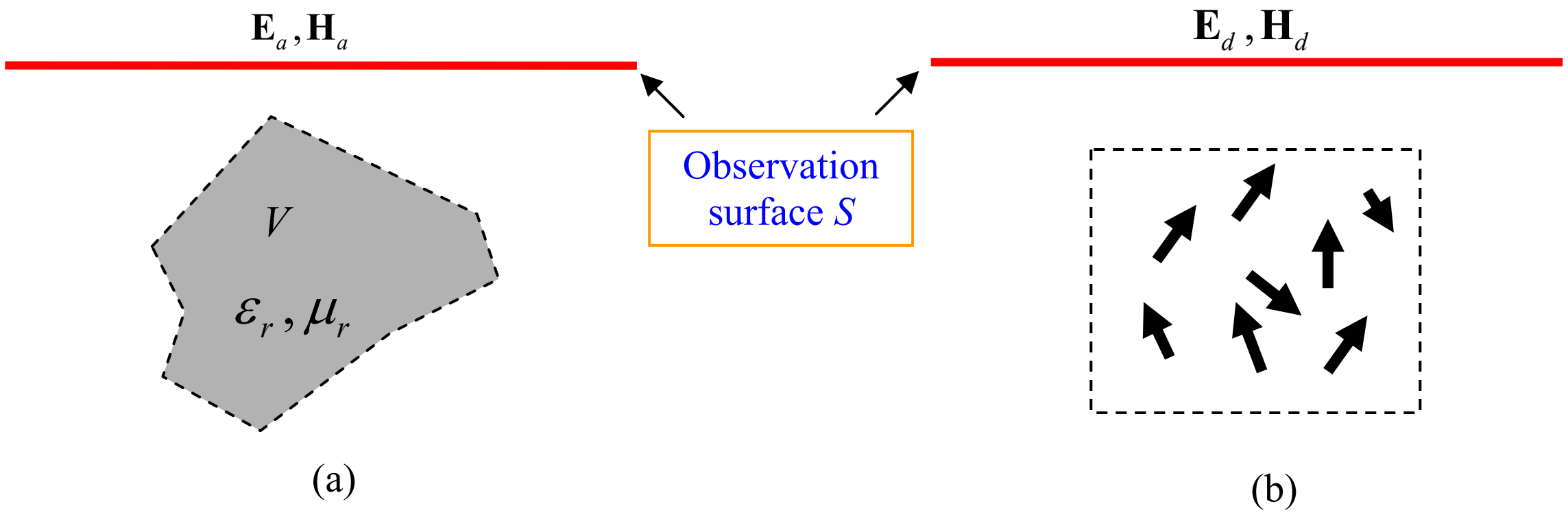

**Fig. 1** (a) Arbitrary antenna (b) An IDM for the antenna in (a)

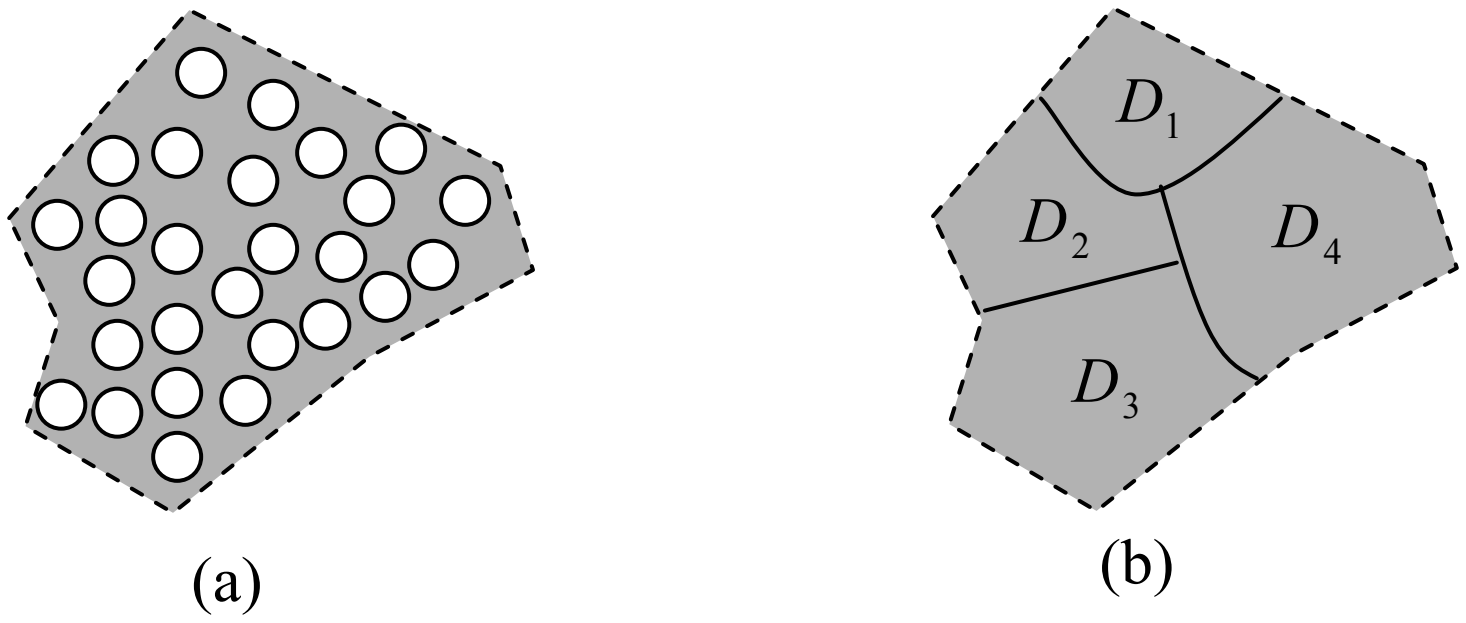

**Fig. 2** (a) Arbitrary antenna modeled by a large number of electric point sources (b) Arbitrary antenna modeled by few number of electric sources, each with finite volume localization.

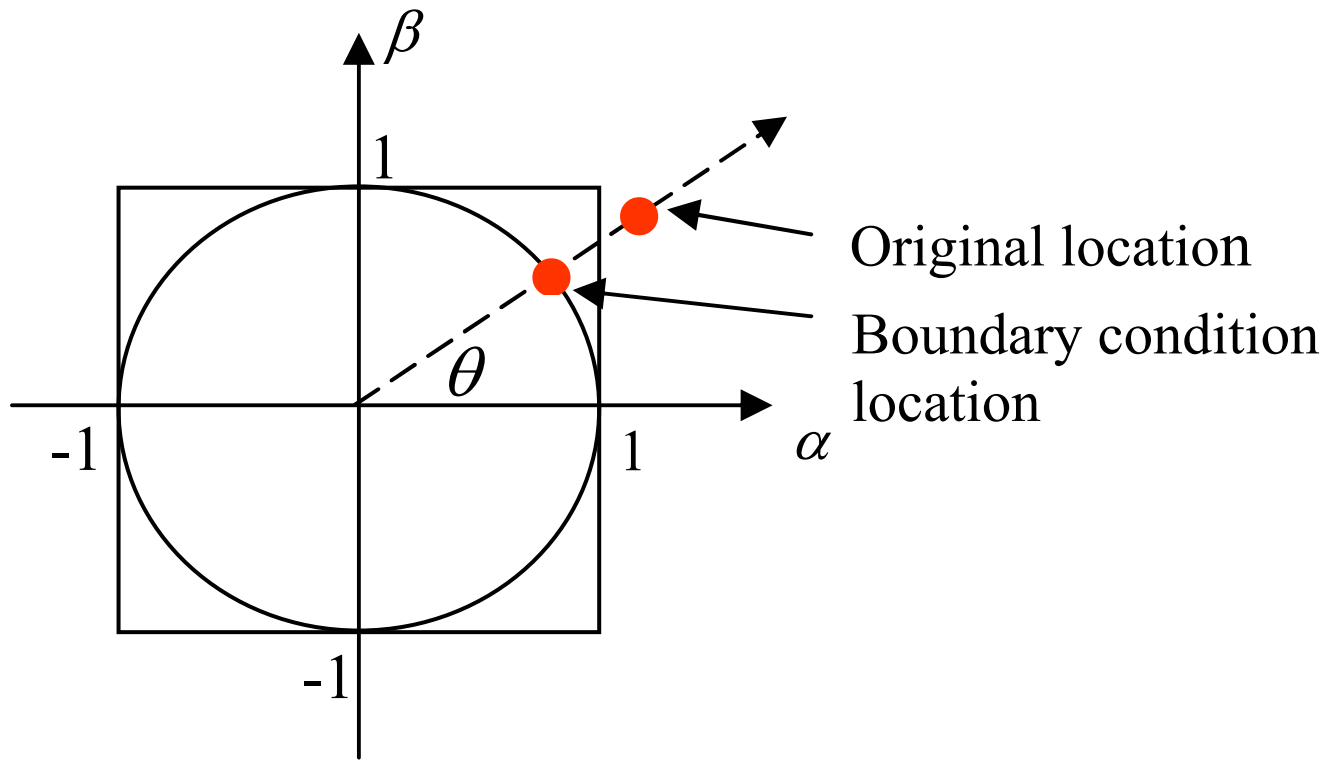

**Fig. 3** The new boundary condition to be used with the QPSO algorithm. The square represents the boundary of un-coupled variables. The circle represents the true boundary of the coupled two direction cosines, $\alpha$ and $\beta$.

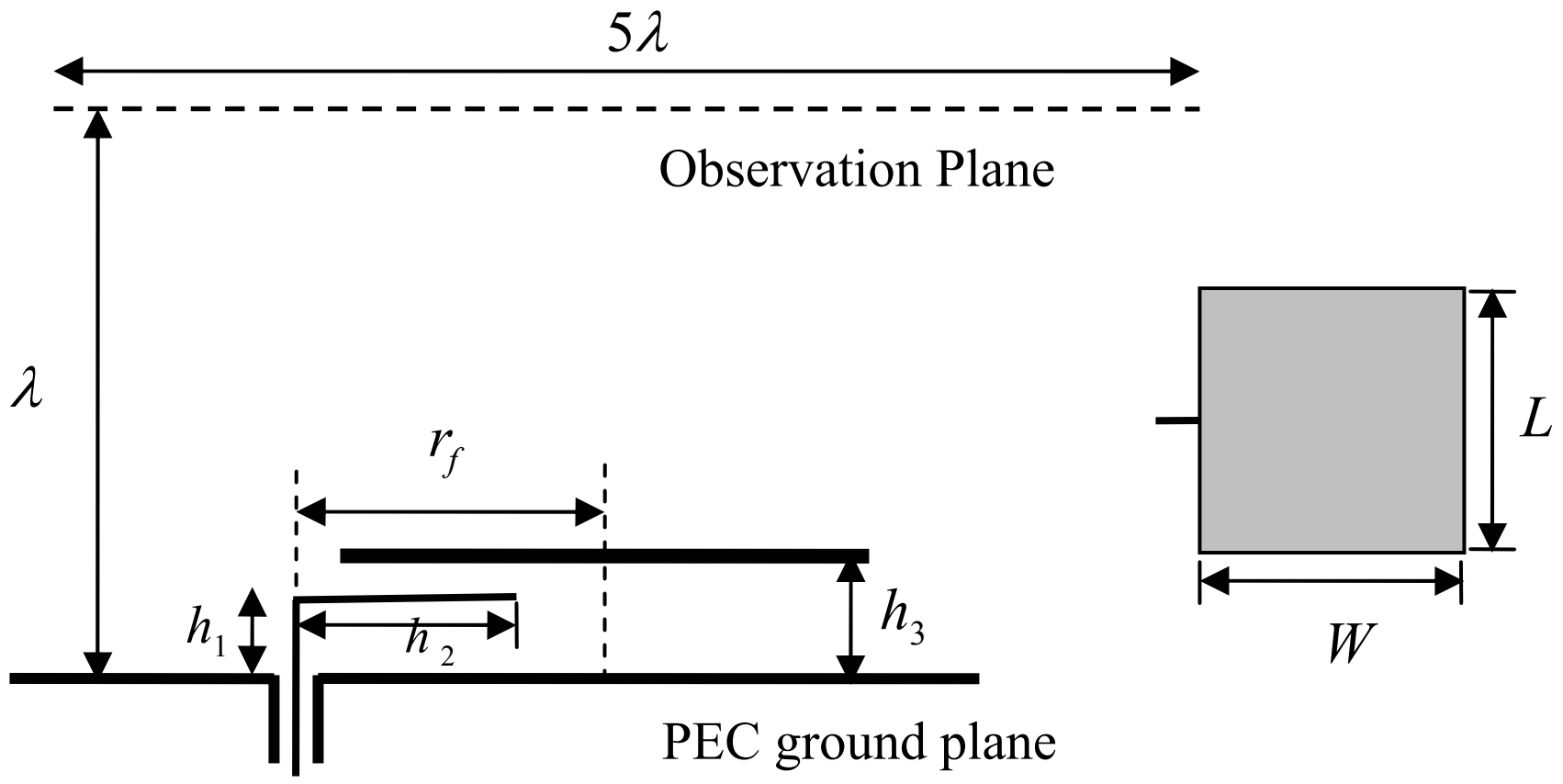

**Fig. 4** Conducting patch excited by L-Shaped coaxial probe. $h_1 = 5\,\mathrm{mm}$, $h_2 = 11.14\,\mathrm{mm}$, $h_3 = 8$ mm, $r_f = 13.84\,\mathrm{mm}$, $W$ = 12.84 mm, and $L$ = 15.44 mm.

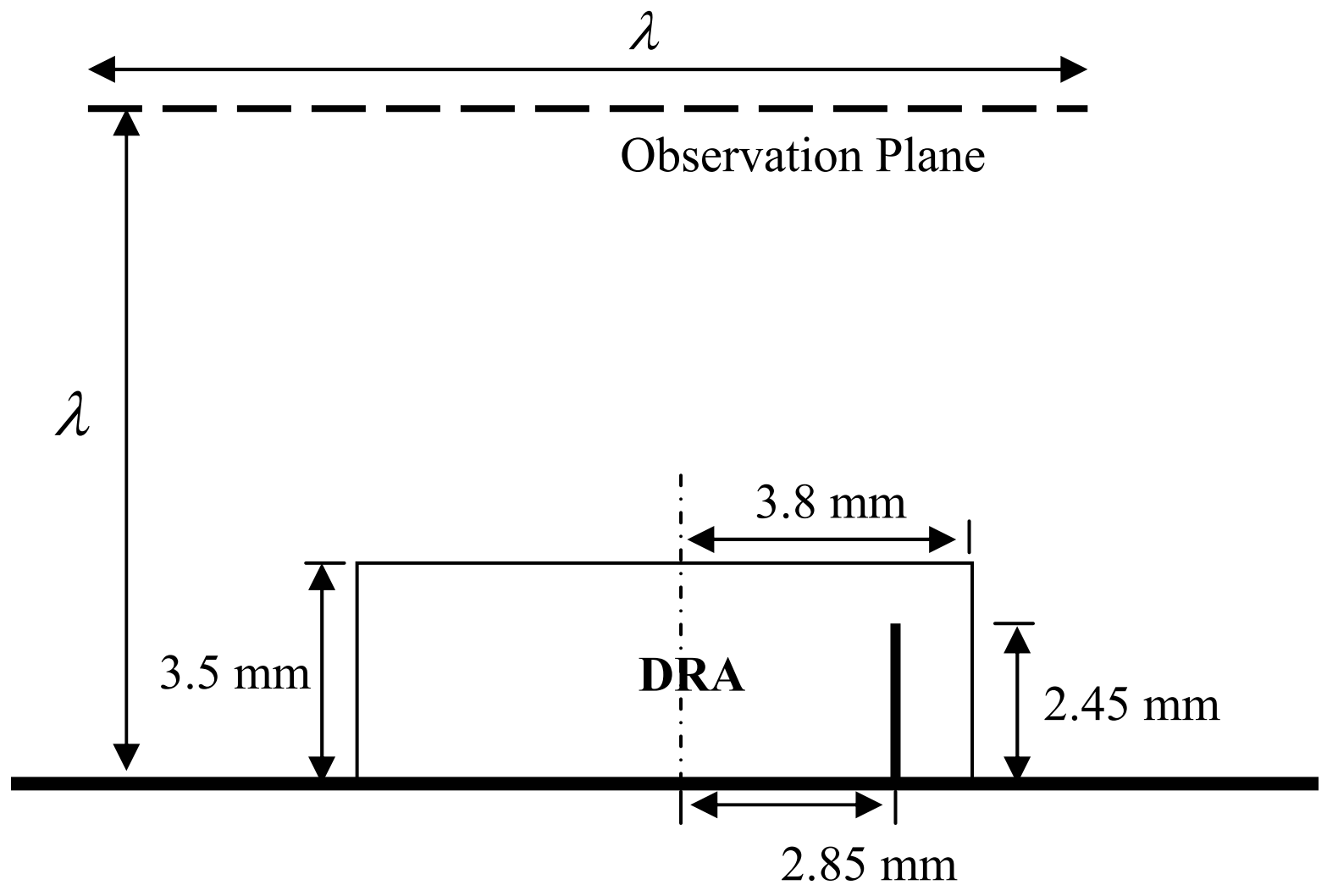

**Fig. 5** Cross sectional view of circular DRA excited by coaxial probe. The relative dielectric constant of the DRA medium is $\varepsilon_r = 10.2$.

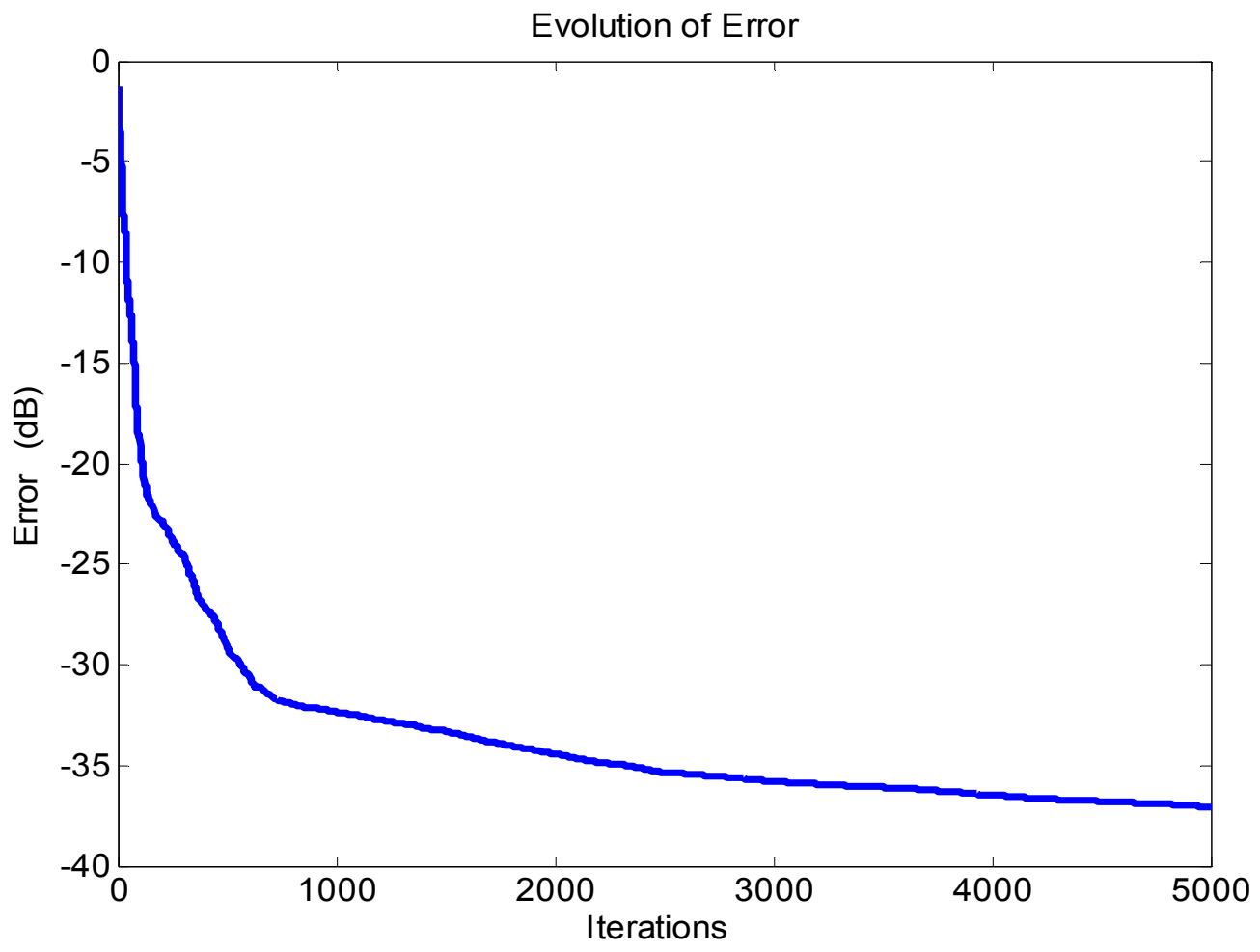

**Fig. 6** Convergence performance of the QDPSO algorithm applied to find the conducting patch's IDM using a set of 10 dipoles with population size of 80 particles and control parameter $g = 3.0$.

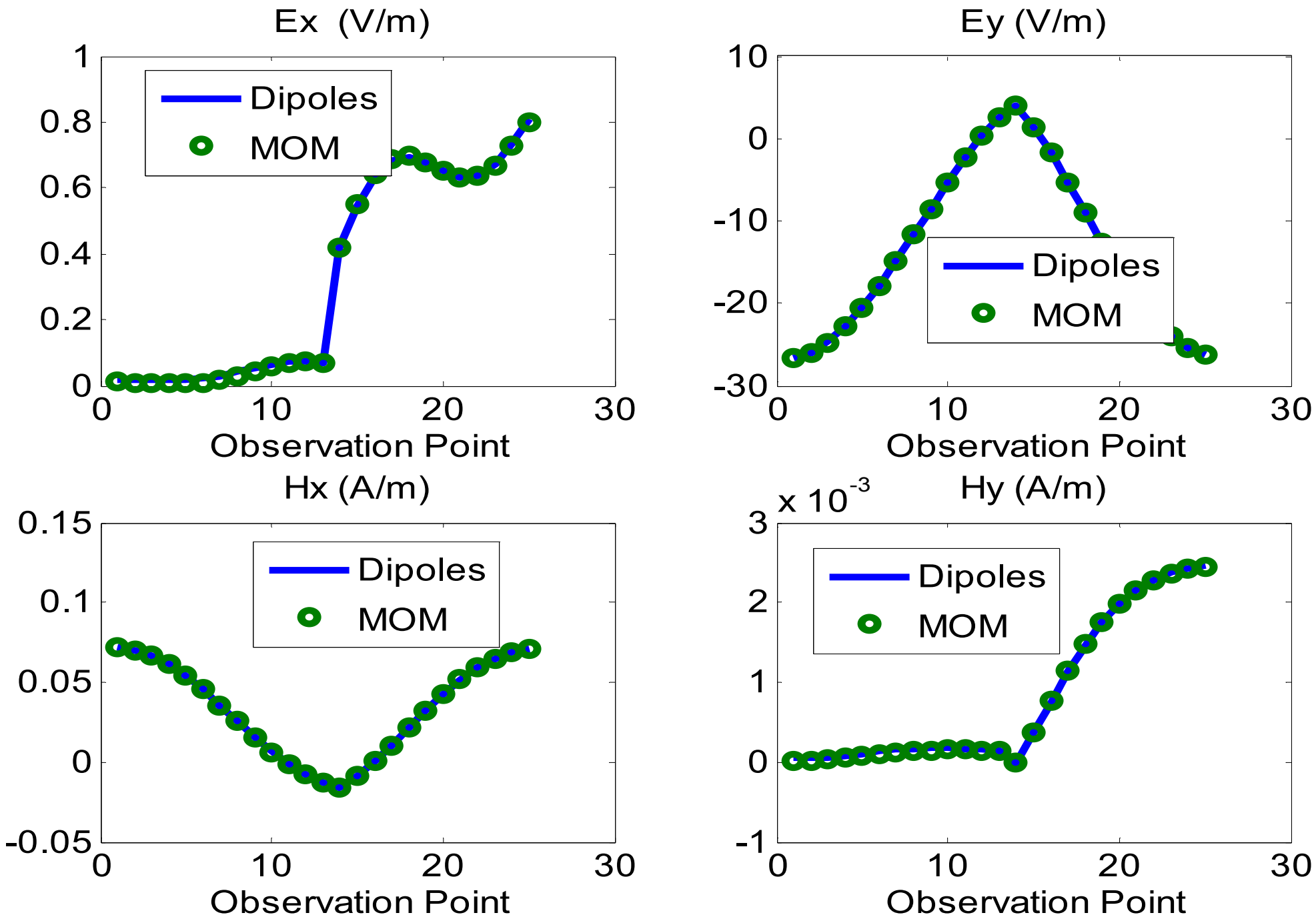

**Fig. 7** Comparison between the MOM and the IDM near-fields for the conducting patch antenna. The tangential electric and magnetic field components are plotted over a line passing in the middle of the observation plane.

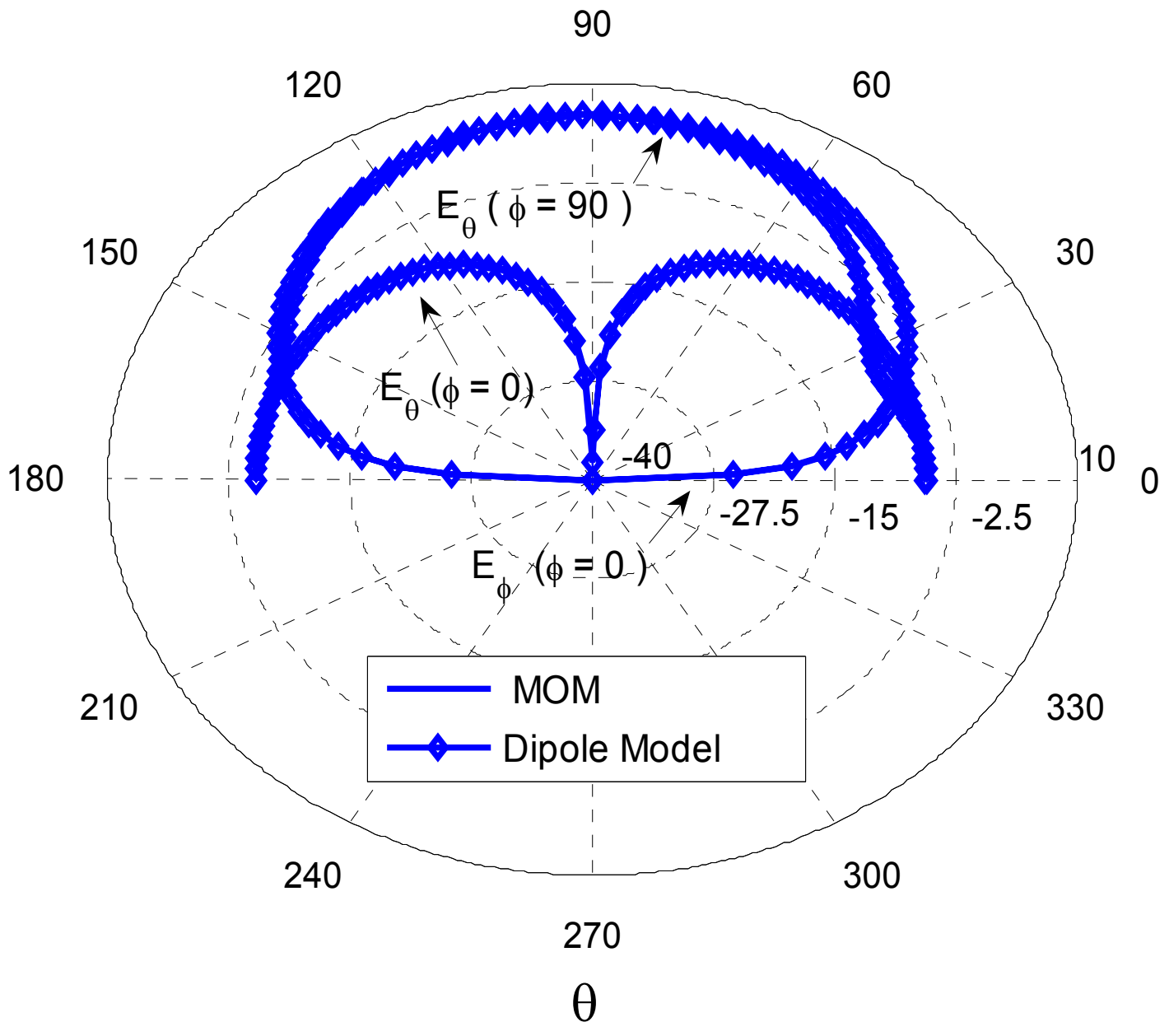

**Fig. 8** Comparison between the MOM and the IDM far-fields of the conducting patch antenna.

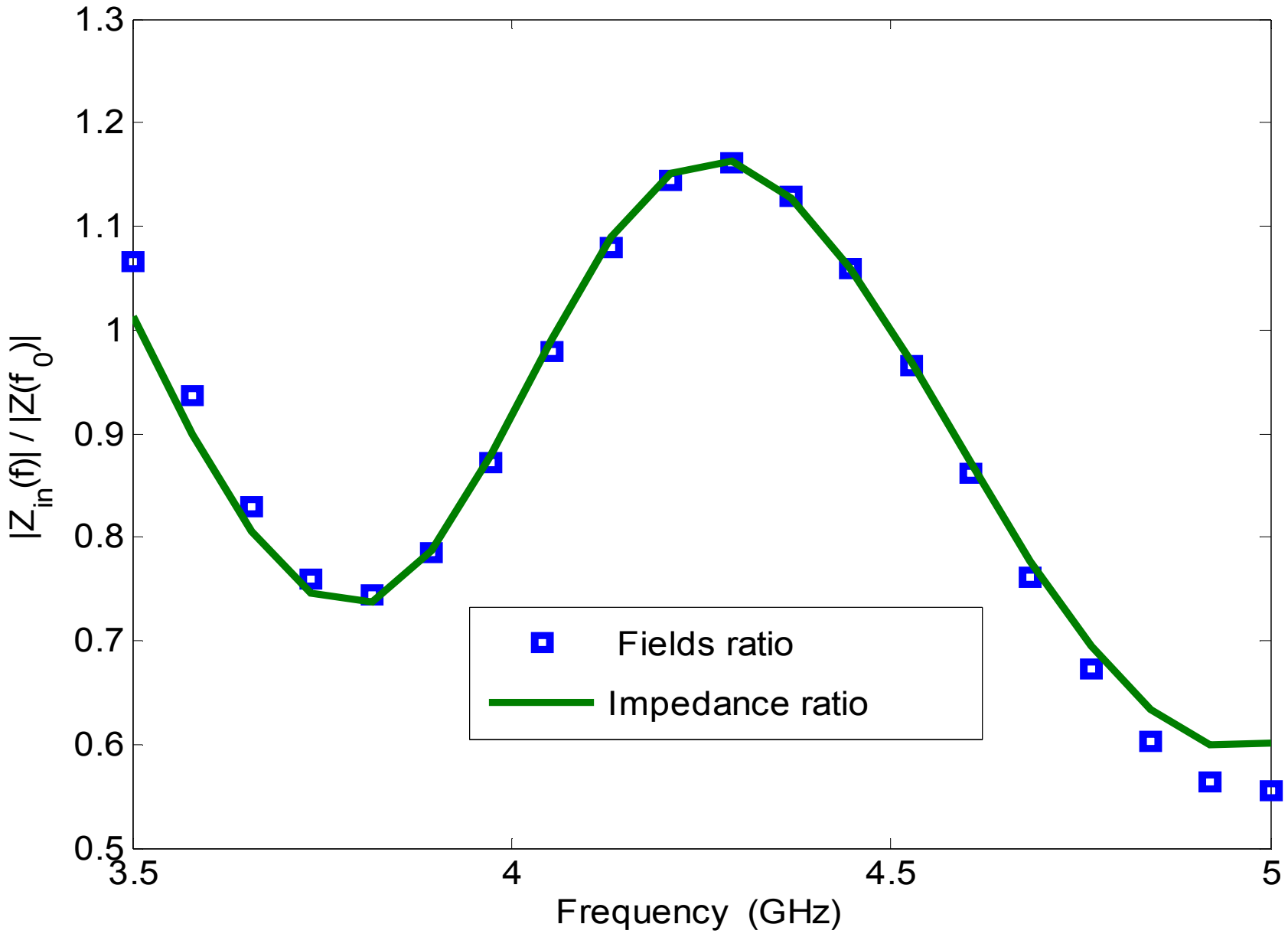

**Fig. 9** The correction factor of the conducting patch wideband IDM.

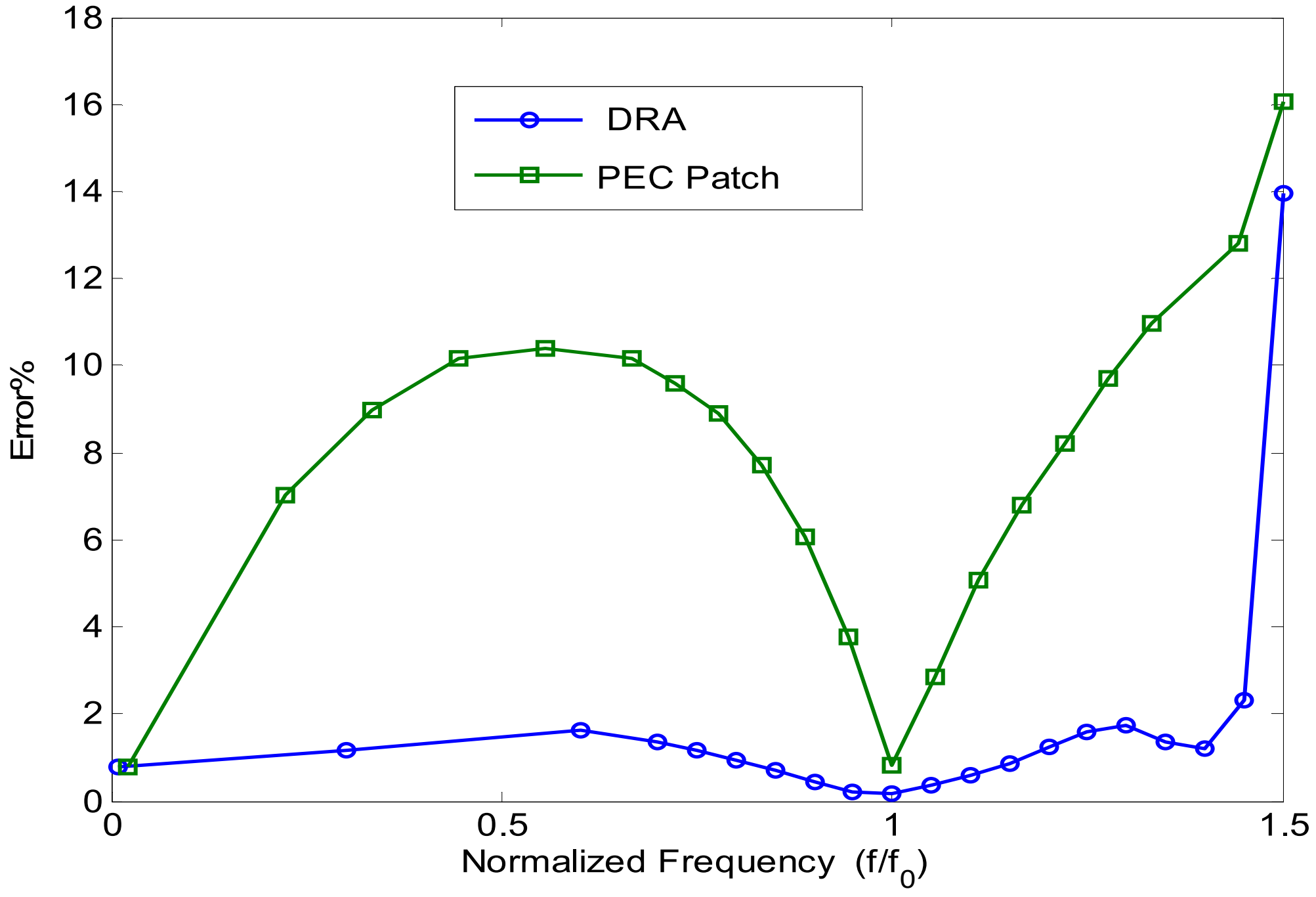

**Fig. 10** Frequency response of the IDM.

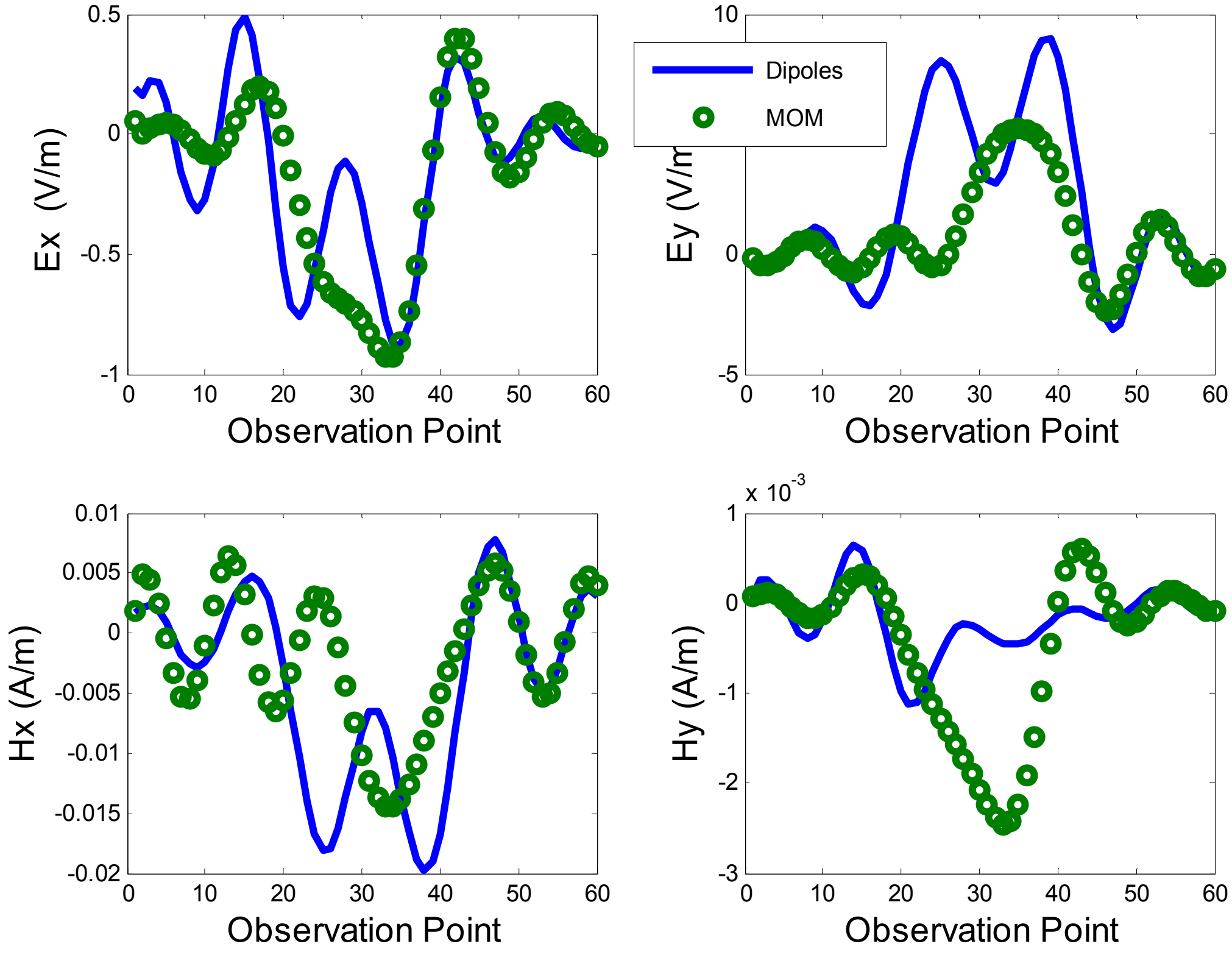

**Fig. 11** Near-field comparison between MOM and the wideband dipole model for the conducting patch antenna obtained at the frequency 6.75 GHz with corresponding error of 16.1%.

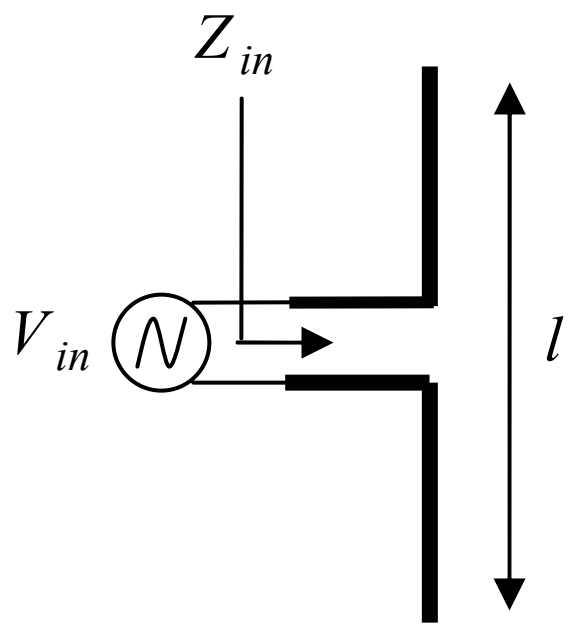

**Fig. 12** Infinitesimal dipole realization by a finite wire structure.

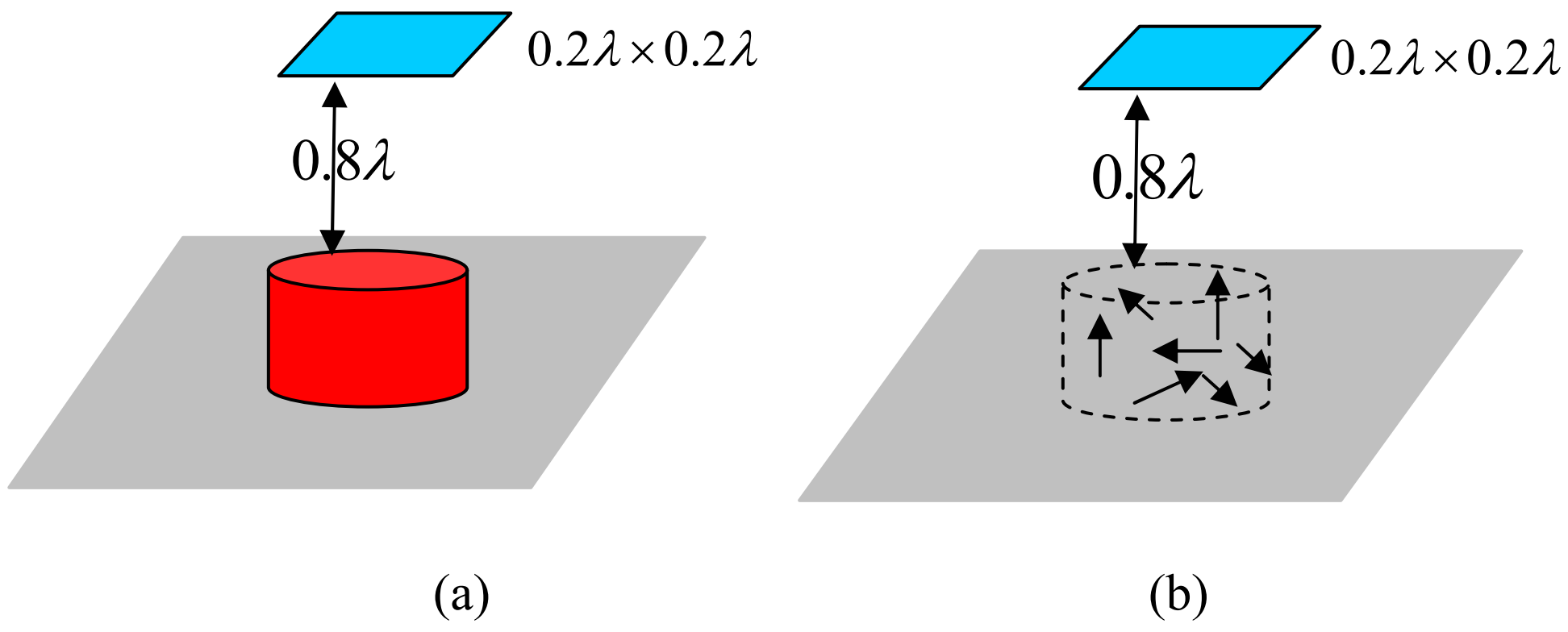

**Fig. 13** (a) Geometry of the DRA close to a PEC square plate. (b) Geometry of the IDM close to the square PEC plate.

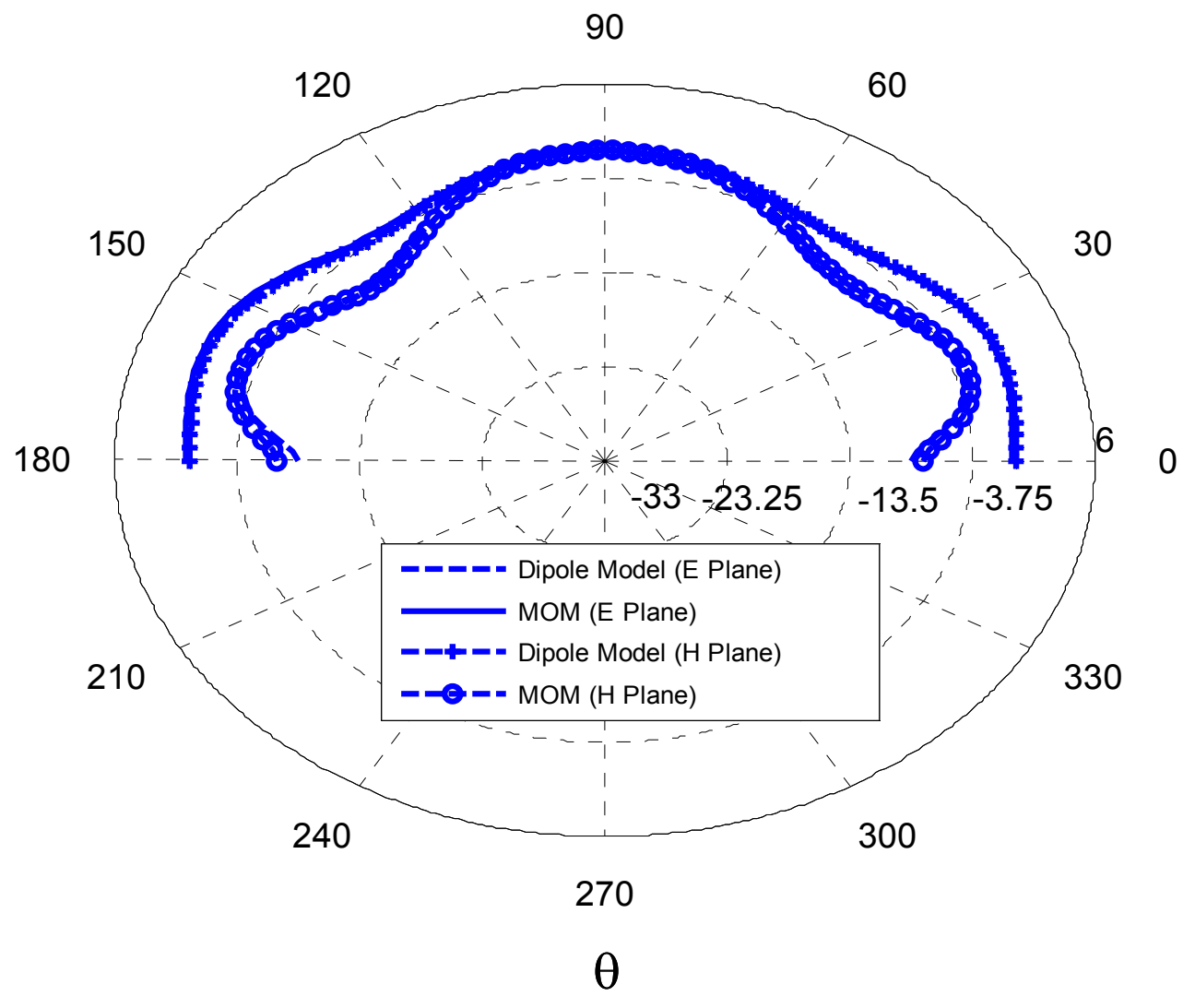

**Fig. 14** Far-Field comparison between the DRA and its IDM in the presence of the conducting plate shown in Fig. 13.

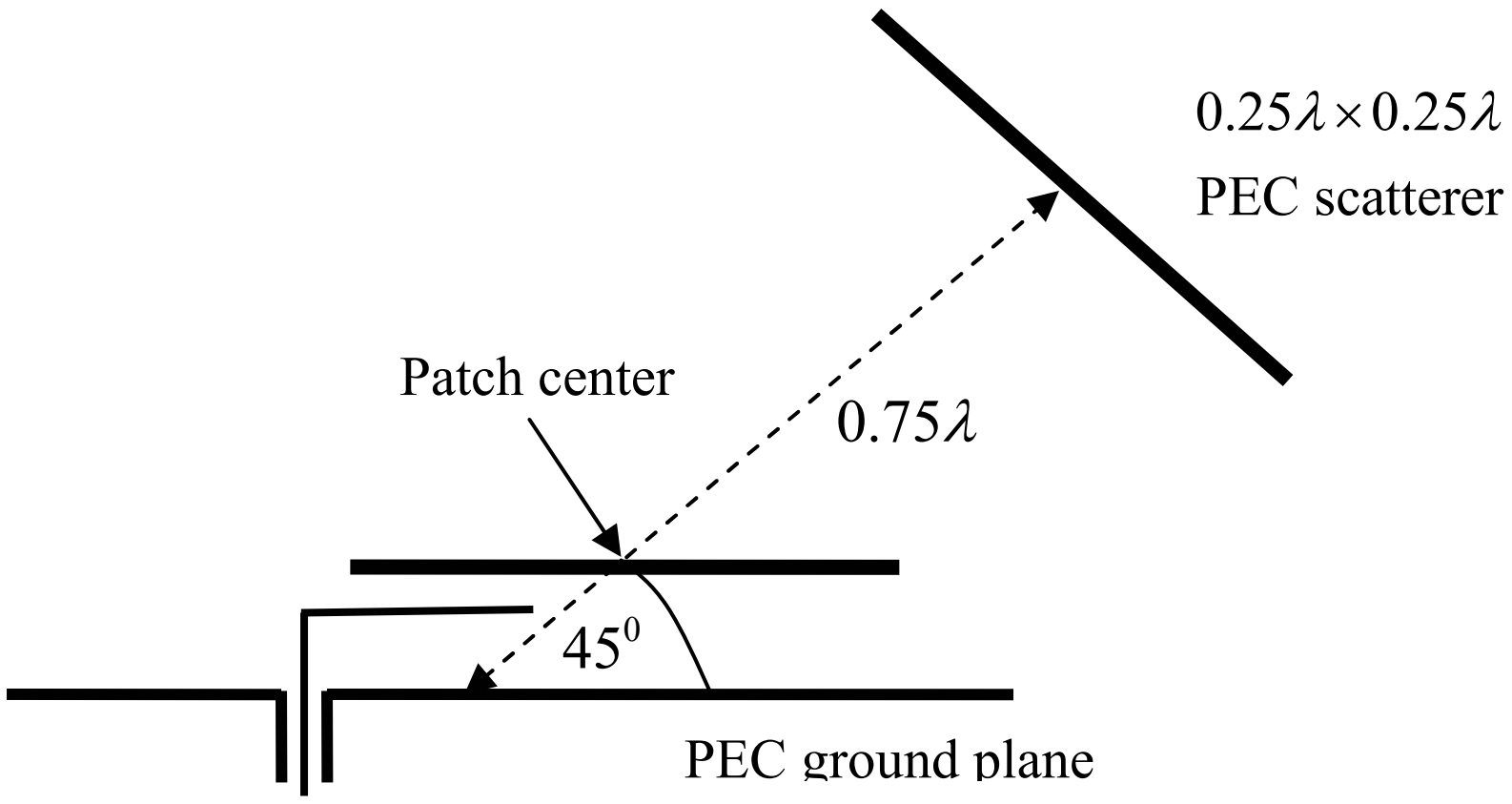

(a)

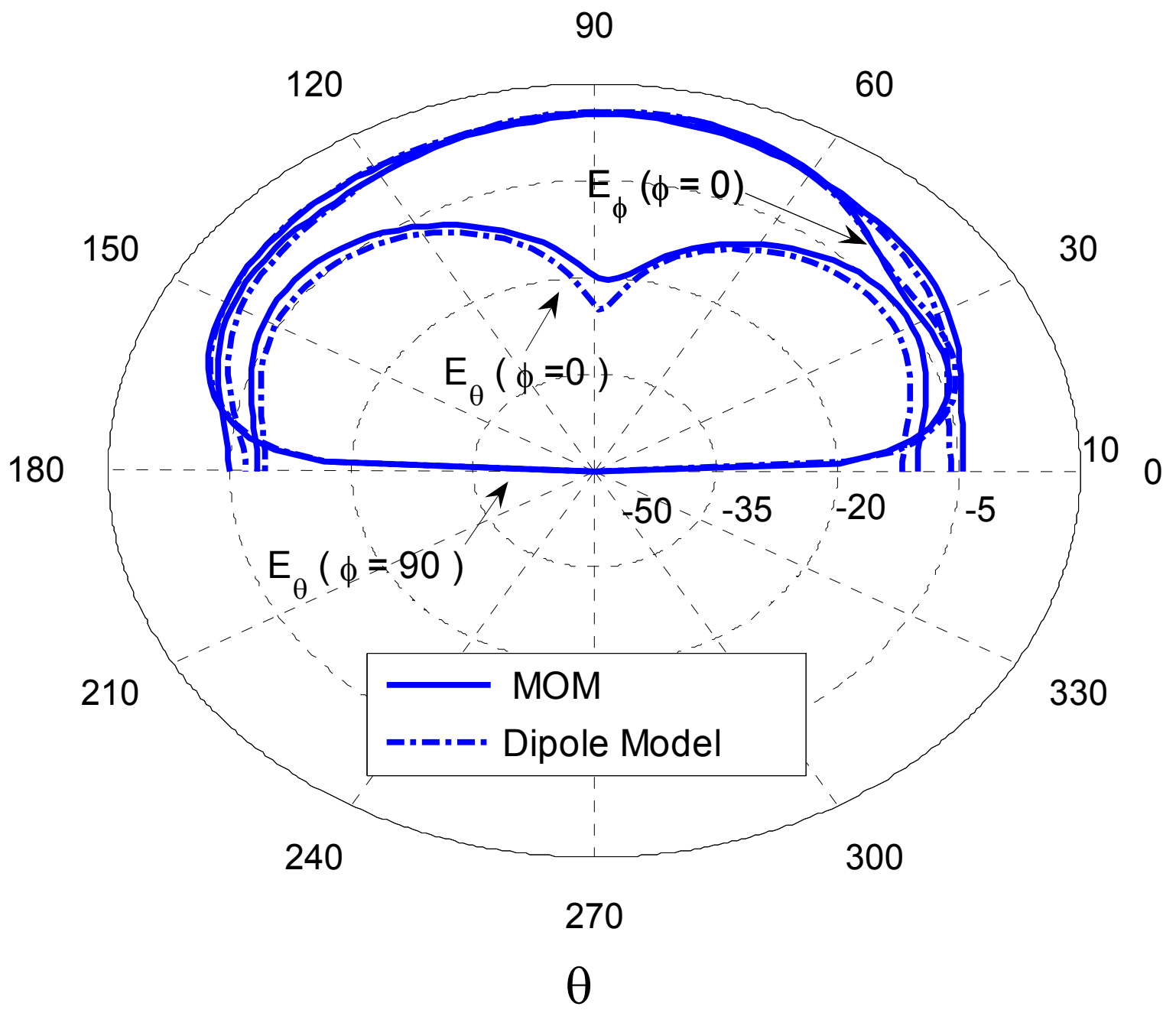

(b)

**Fig. 15** (a) Geometry of the conducting patch antenna close to a rotated PEC square plate. (b) Far-field comparison between the IDM prediction and the original problem solution.

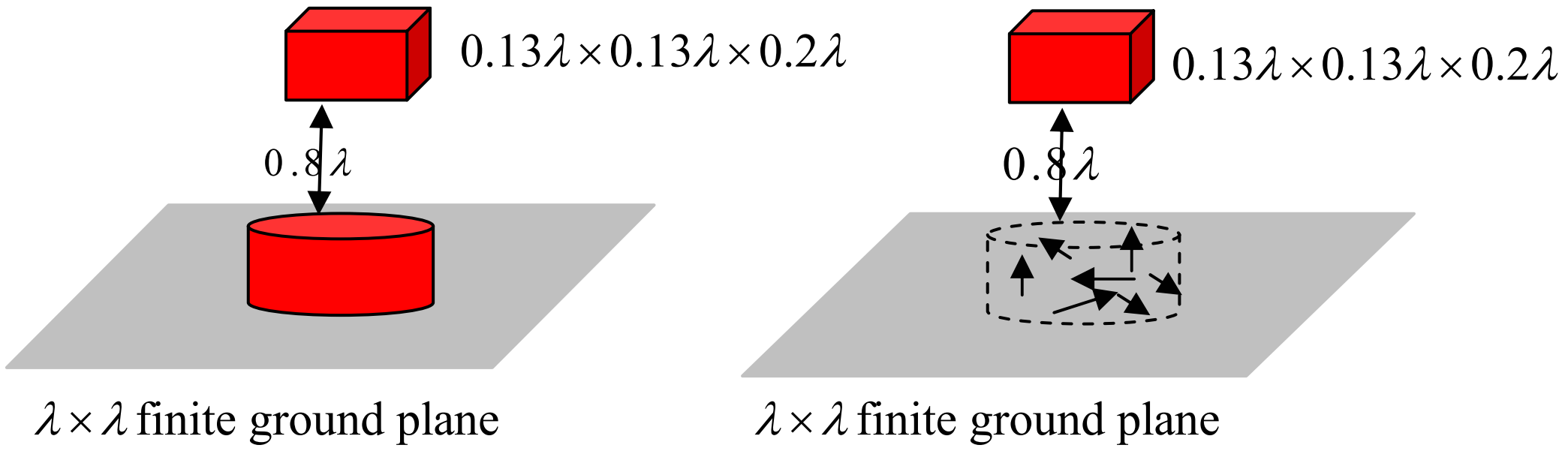

(a) (b)

**Fig. 16** (a) Geometry of the DRA close to a dielectric cube with $\varepsilon_r = 3.0$ . (b) Geometry of the IDM close to a dielectric cube.

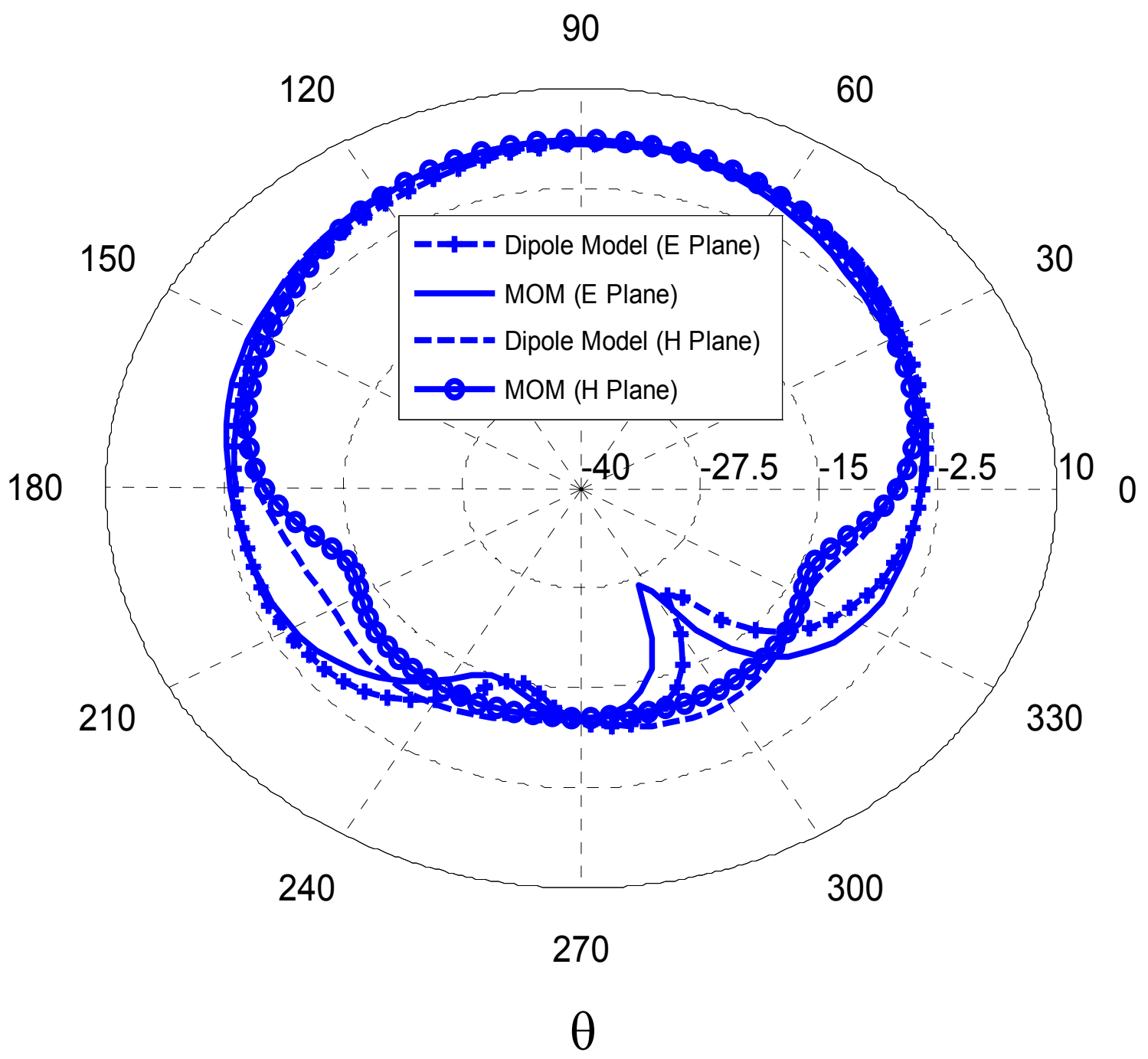

**Fig. 17** Far-Field comparison between the DRA and its IDM in the presence of finite ground plane as shown in Fig. 17.

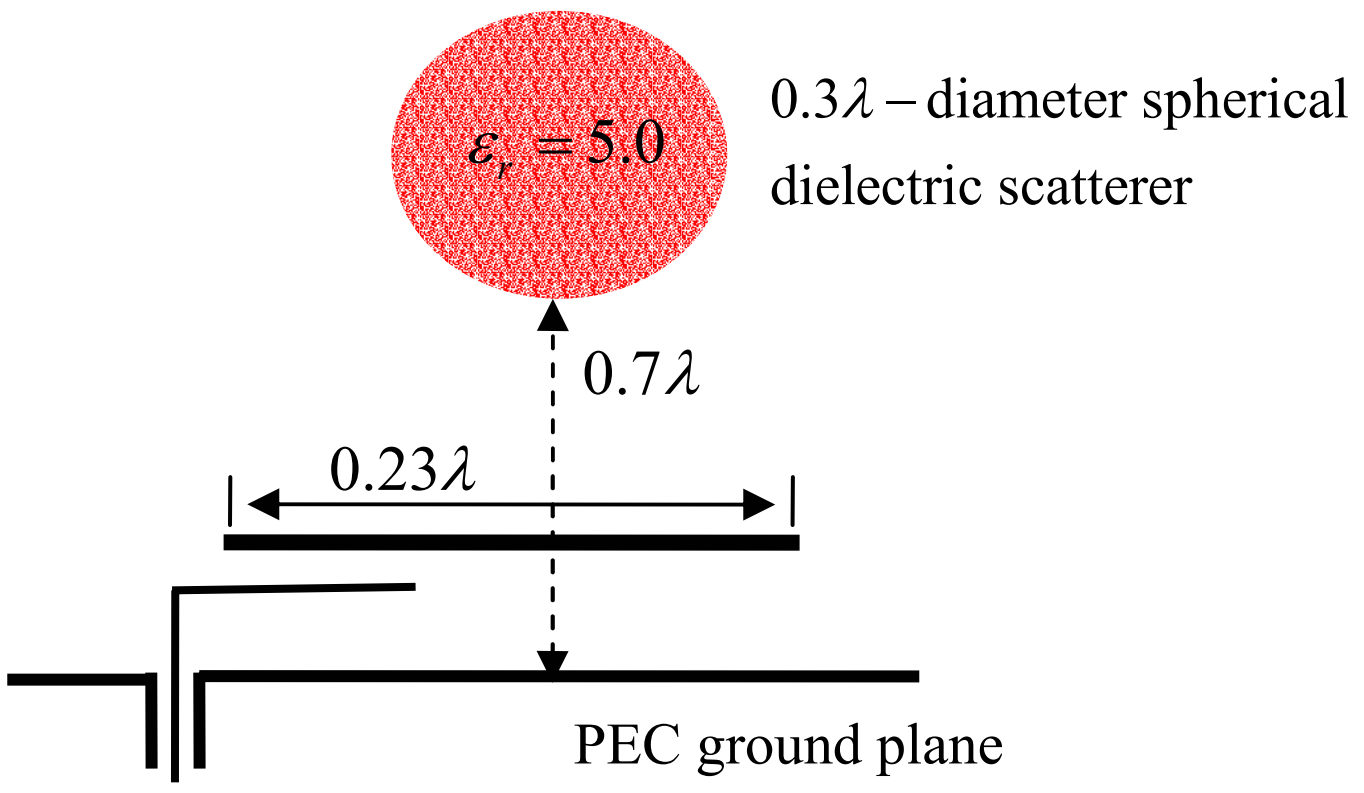

(a)

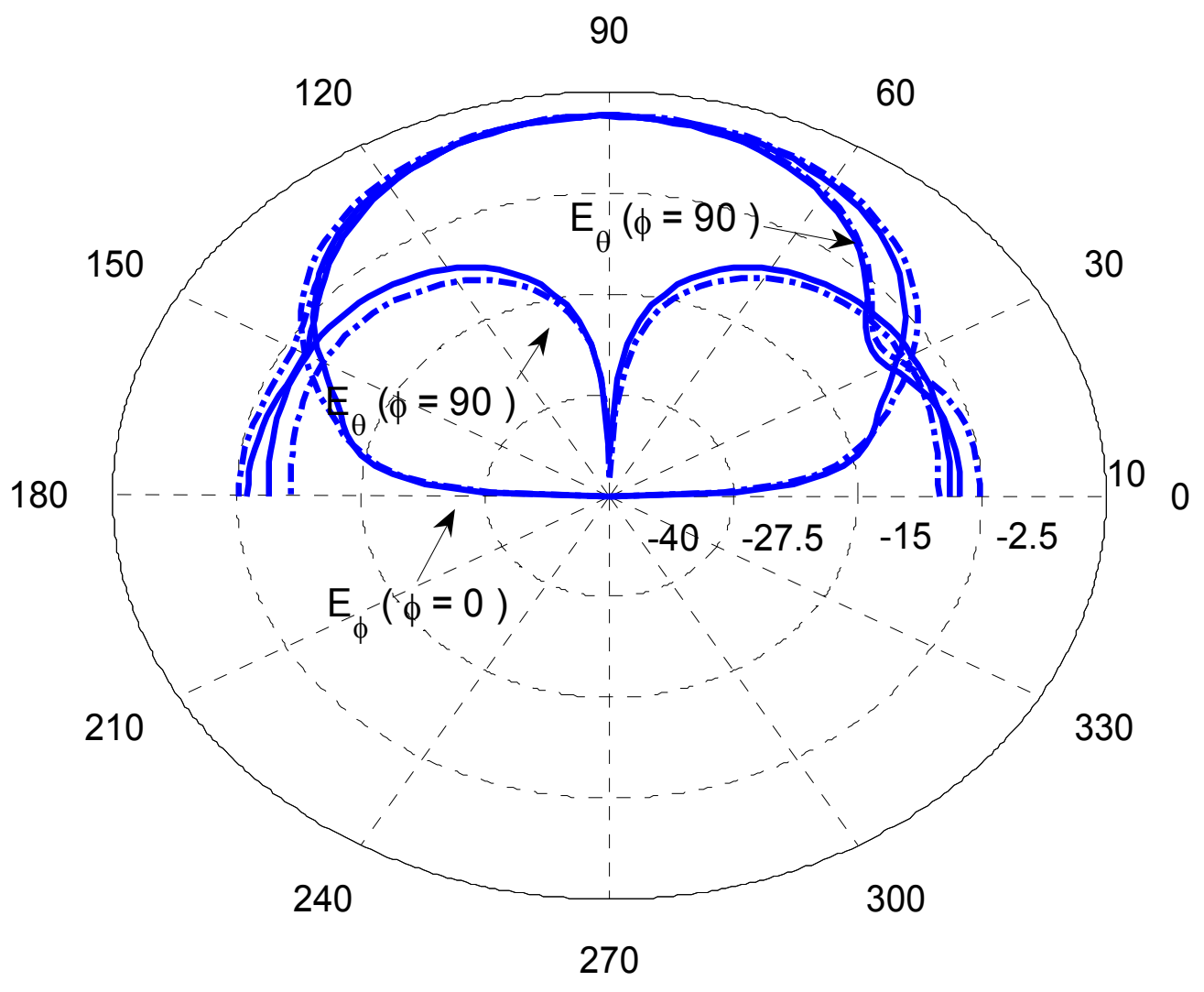

(b)

**Fig. 18** (a) Geometry of the conducting patch antenna close to a spherical dielectric object. (b) Far-field comparison between the IDM prediction and the original problem solution.

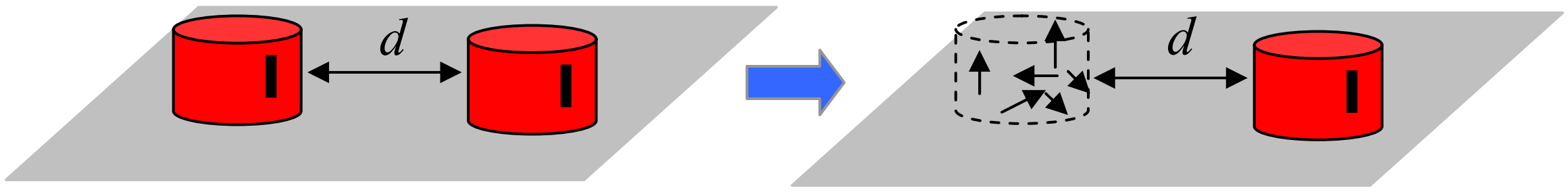

(a)

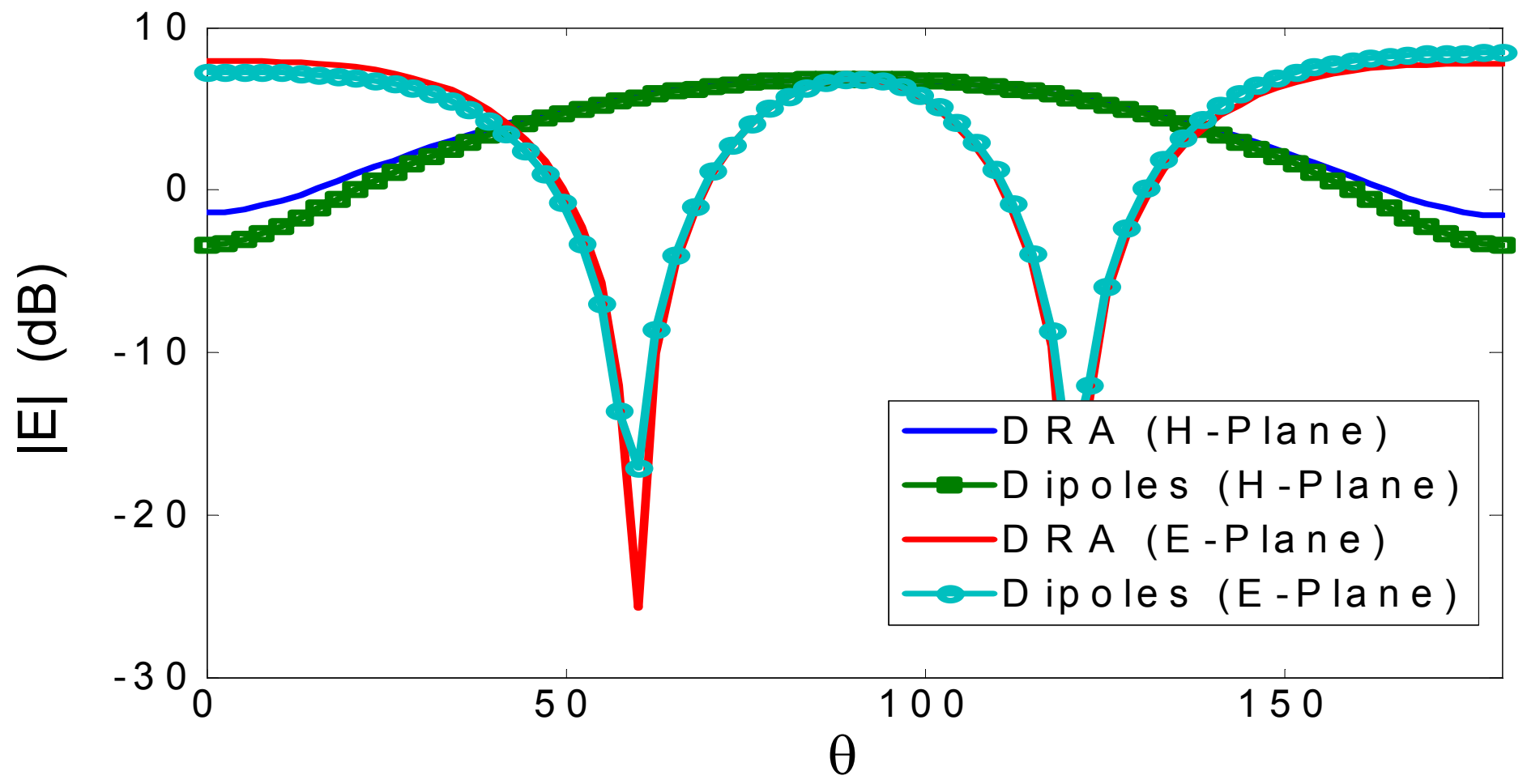

(b)

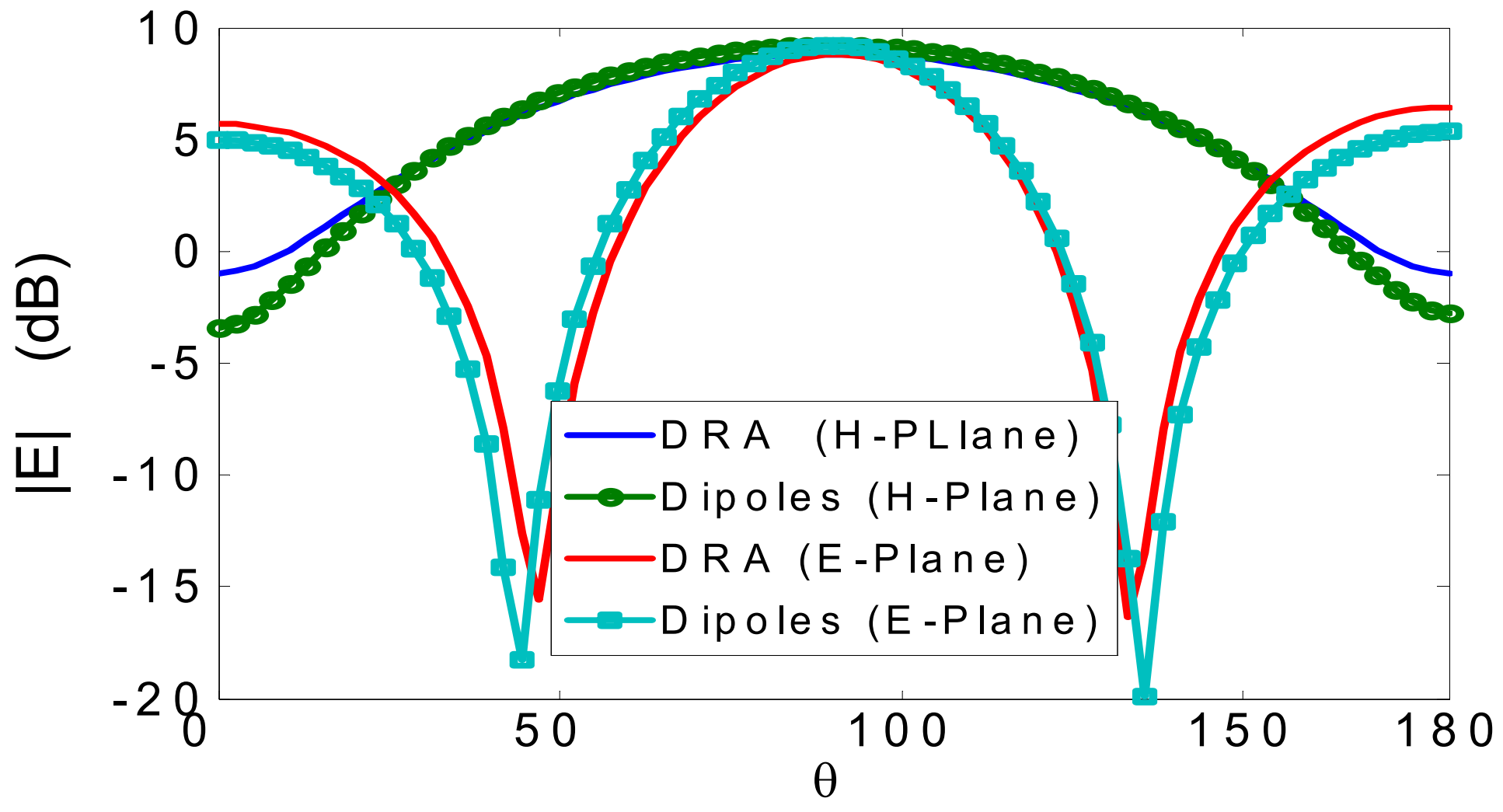

(c)

**Fig. 19** (a) E-plane 2-element DRA array where one DRA is replaced by its IDM (b) Far-field comparison for $d = \lambda$ (c) Far-field comparison for $d = 0.67\lambda$ .

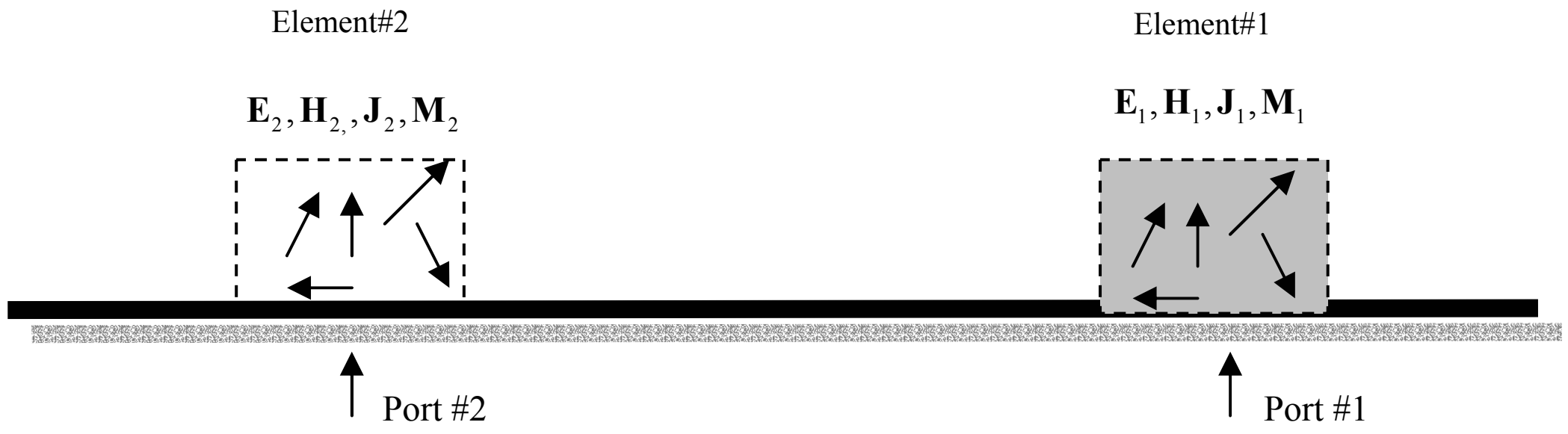

**Fig. 20** Schematic diagram for the interaction between two arbitrary antenna elements.

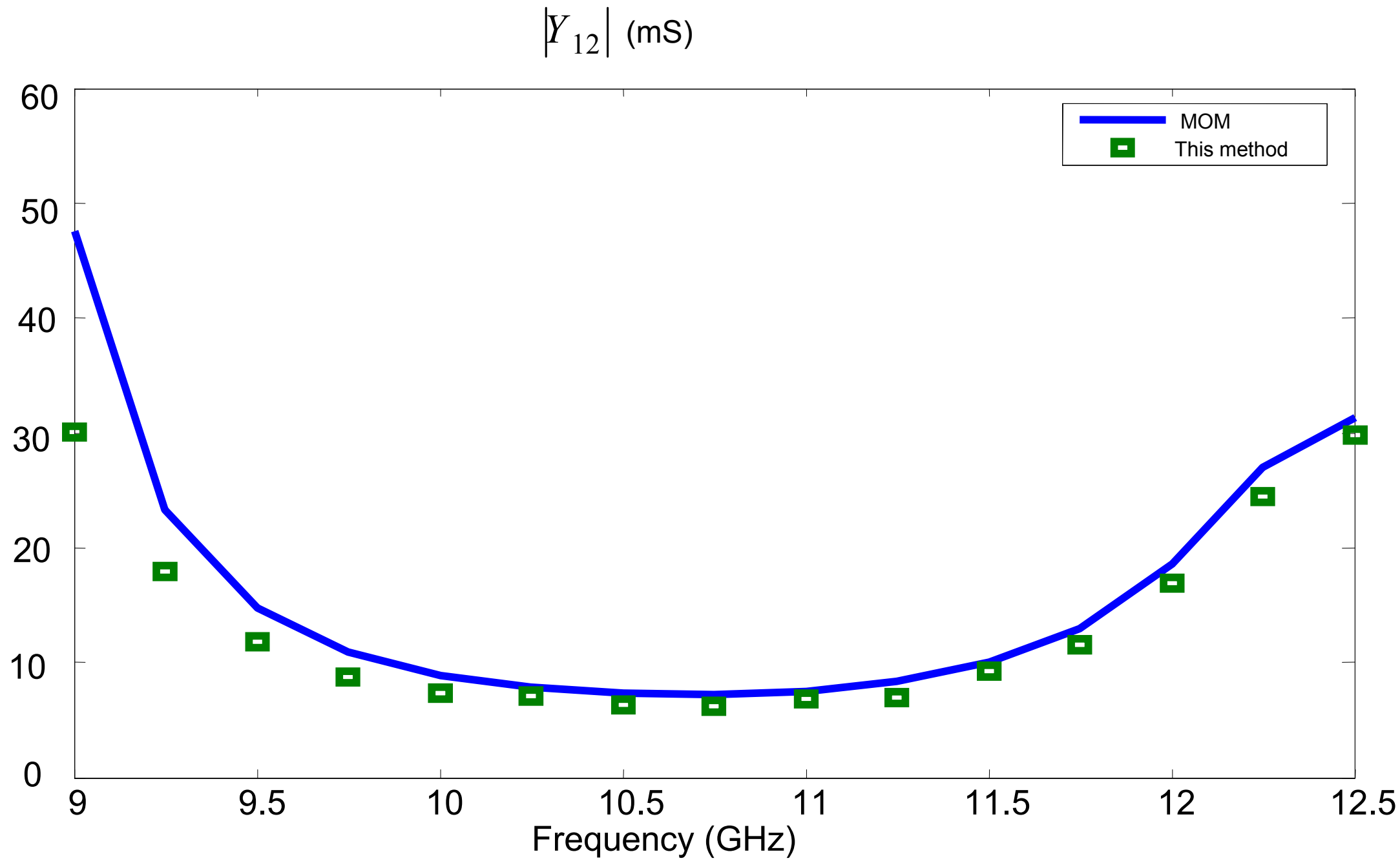

**Fig. 21** Comparison between the MOM solution and the proposed method for the calculation of mutual coupling between 2-element array, where each element is identical to the geometry in Fig. 5 separated by a distance of $0.5\lambda$ .

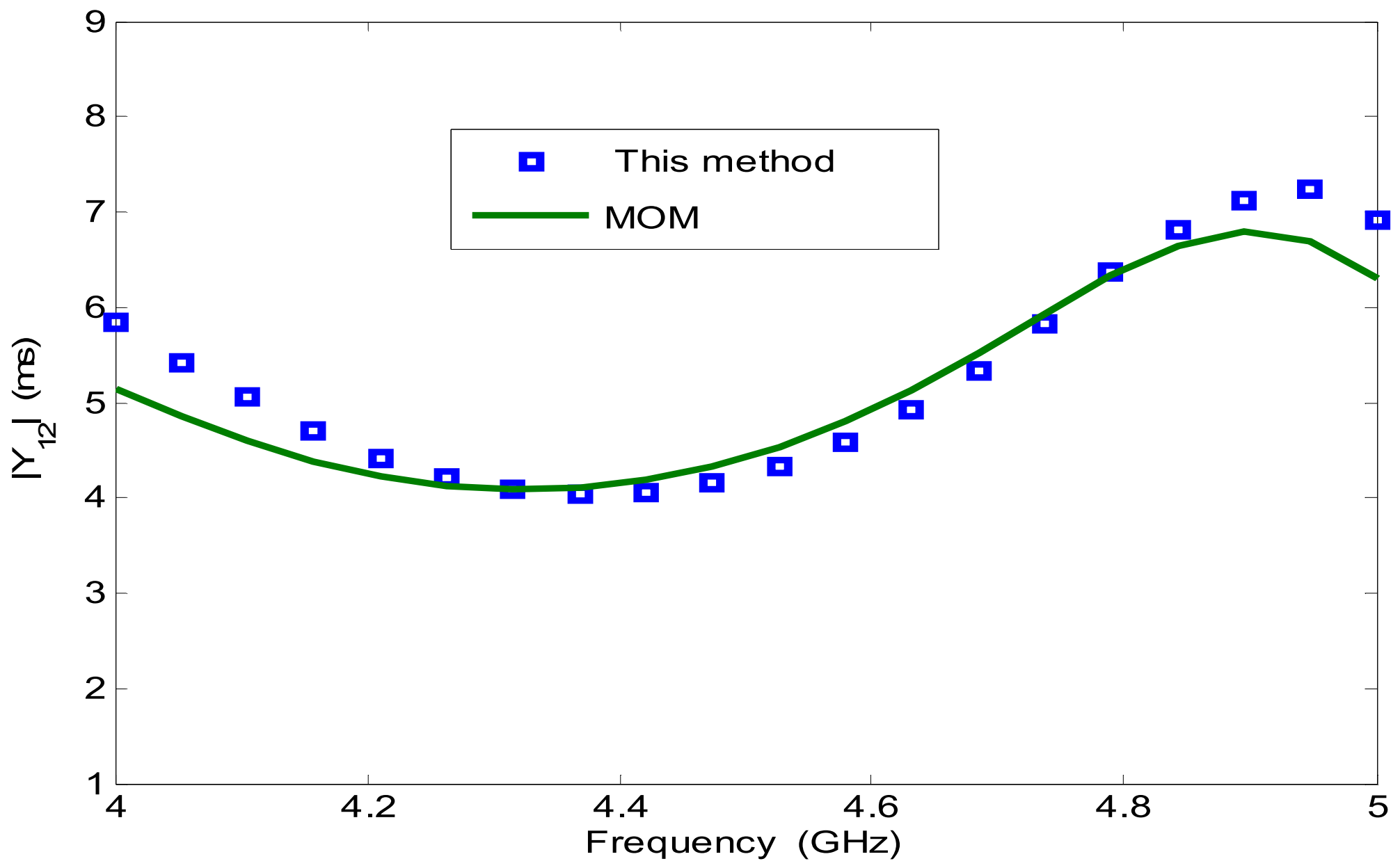

**Fig. 22** Comparison between the MOM solution and the proposed method for the calculation of mutual coupling between two array elements identical to the geometry in Fig. 4 separated by a distance of $0.5\lambda$ .